\documentclass[pra,twocolumn,superscriptaddress,a4paper,english,longbibliography]{revtex4-2}

\usepackage{lineno}
\usepackage{times}
\usepackage{color}
\usepackage[colorlinks=true,urlcolor=blue,citecolor=blue]{hyperref}
\usepackage{url}
\usepackage{breakurl}
\usepackage{soul}
\usepackage{graphicx}
\usepackage{amsmath}
\usepackage{subfigure}
\usepackage{xcolor}
\usepackage{amssymb}
\usepackage{hyperref}
\DeclareGraphicsExtensions{.png,.eps}
\usepackage{float}

\usepackage{appendix}
\usepackage{algorithm}
\usepackage{algpseudocode}

\newcommand {\R}{\textcolor {black}}
\newcommand {\RR}{\textcolor {black}}

\newcommand {\RRR}{\textcolor {black}}
\usepackage{xcolor}
\usepackage[urlcolor=blue]{hyperref}      
\hypersetup{
    colorlinks = true,                    
    citecolor = {blue},
    linkcolor = {purple},
           }

\begin{document}


\title{Quantum compiling with variational instruction set for accurate and fast quantum computing}

\author{Ying Lu}
\affiliation{Department of Physics, Capital Normal University, Beijing 100048, China}
\author{Peng-Fei Zhou}
\affiliation{Department of Physics, Capital Normal University, Beijing 100048, China}
\author{Shao-Ming Fei} \email[Corresponding author. Email: ] {feishm@cnu.edu.cn}
\affiliation{School of Mathematical Sciences, Capital Normal University, Beijing 100048, China}
\affiliation{Max-Planck-Institute for Mathematics in the Sciences, 04103, Leipzig, Germany}
\author{Shi-Ju Ran} \email[Corresponding author. Email: ] {sjran@cnu.edu.cn}
\affiliation{Department of Physics, Capital Normal University, Beijing 100048, China}
\date{\today}

\begin{abstract}
	The quantum instruction set (QIS) is defined as the quantum gates that are physically realizable by controlling the qubits in a quantum hardware. Compiling quantum circuits into the product of the gates in a properly-defined QIS is a fundamental step in quantum computing. We here propose the \R{quantum variational instruction set (QuVIS)} formed by flexibly-designed multi-qubit gates for higher speed and accuracy of quantum computing. The controlling of qubits for realizing the gates in a QuVIS are variationally achieved using the fine-grained time optimization algorithm. Significant reductions on both the error accumulation and time cost are demonstrated in realizing the swaps of multiple qubits and quantum Fourier transformations, compared with the compiling by the standard QIS such as \RR{the quantum microinstruction set} (QuMIS, formed by several one- and two-qubit gates including the one-qubit rotations and controlled-NOT gate). With the same requirement on quantum hardware, the time cost by \R{QuVIS} is reduced to be less than one half of that by QuMIS. Simultaneously, the error is suppressed algebraically as the depth of the compiled circuit is reduced. As a general compiling approach with high flexibility and efficiency, \R{QuVIS} can be defined for different quantum circuits and adapt to the quantum hardware with different interactions.
\end{abstract}

\maketitle

\section{Introduction}


Efficient compiling of quantum algorithms to physically-executable forms belongs to the fundamental issues of quantum computing. A widely-recognized compiling way is to transform the circuit into the product of executable elementary gates, which are named as quantum instruction set (QIS)~\cite{green2013quipper,wecker2014liqui,javadiabhari2015scaffcc, chong2017programming, haner2018software}. A QIS should be constructed according to the fundamental physical mechanism of quantum hardware. For instance, a superconducting quantum computer can adopt the \RR{quantum microinstruction set} (QuMIS)~\cite{fu2017experimental} as the instructive set that is formed by several one- and two-qubits gates including the one-qubit rotations and controlled-NOT (CNOT). For the quantum photonic circuits, the elementary gates represent certain basic operations on single photons~\cite{OFV09Qphoto, AW12Qphoto}. The efficiency of compiling a given quantum algorithm with a chosen QIS can be characterized by the complexity (e.g., depth) of the compiled circuit.

A typical way of realizing the elementary gates in a QIS is by controlling the \RR{dynamics of quantum hardware. Different quantum platforms are usually described by different controlling process. For instance, superconducting circuits employ microwave-pulse techniques to manipulate the qubits~\cite{krantz2019quantum, PhysRevApplied.6.064007} through, e.g., cross resonance~\cite{PhysRevLett.107.080502}, parametric modulation~\cite{doi:10.1126/sciadv.aao3603}, and etc.. Another typical quantum hardware is the nuclear magnetic resonance (NMR) systems~\cite{cory1998nuclear, jones1998quantum, jones1998implementation, vandersypen2001experimental, RevModPhys.76.1037, bian2017universal, chuang1998experimental, jones1998fast, zhang2002realization}. For such systems, a key issue of realizing quantum circuits or algorithms, such as Shor’s factoring algorithm~\cite{doi:10.1137/S0036144598347011} and Harrow-Hassidim-Lloyd related algorithms~\cite{PhysRevLett.103.150502}, is to determine the tunable time-dependent pulses.} The efficiency can be characterized by the time cost of the controlling process.

For the two-qubit gates, such as CNOT and swap gates, the optimal time cost has theoretically-given bounds~\cite{khaneja2001time, li2013time, sun2020time}. For the $N$-qubit gates with $N>2$, such bounds are not rigorously given in most cases, and variational methods including the machine learning (ML) techniques have recently been adopted in such optimal-control problems~\cite{PhysRevA.52.R891, PhysRevA.95.042318, PhysRevX.7.021027, PhysRevA.98.052125, wu2020end, PhysRevResearch.3.023092, magann2021pulses, PhysRevA.103.022613, PhysRevA.103.012404, lu2021preparation, PhysRevResearch.4.L012029}. Besides, quantum many-body systems have also been used to implement the measurement-based quantum computation~\cite{PhysRevLett.101.010502.2008, MBQCrev2009, PhysRevLett.108.240505.2012, PhysRevA.85.010304.2012, Darmawan_2012, PhysRevA.92.012310.2015, PhysRevA.96.032317.2017}. However, most conventional methods concern the controlling of a few qubits. The utilizations of the many-body dynamics for quantum computing~\cite{PhysRevX.7.021027, PhysRevResearch.3.023092, magann2021pulses, lu2021preparation, PhysRevA.105.022618, leng_differentiable_2022, huang_time-optimal_2022} are much less explored due to the exponentially-high complexity.

For all known quantum computing platforms, \RR{noise is} inevitable and will induce computational errors that make the results unstable or unreliable. One way of fighting against errors is the error correction codes~\cite{LB13QECbook}, such as Calderbank-Shor-Steane codes~\cite{PhysRevA.54.1098}, Reed-Muller quantum codes~\cite{771249}, and Toric codes~\cite{K03Toric}.  \RR{However, the implementation of quantum error correction codes will significantly increase not only the number of qubits but also the complexity of circuits. This issue is particularly important in the noisy intermediate-scale quantum (NISQ) era, where the number of available qubits and the connectivity among them are limited. Besides, noise} will also lead to decoherence, meaning that the qubits will gradually become less entangled, losing the supremacy over classical computing. Prolonging the coherence time and reducing the time cost so that the quantum computing tasks are executed within the coherence duration belong to the significant and challenging issues for quantum computing in the NISQ era (see, e.g., Refs.~[\onlinecite{S95decoh, Science1995, PhysRevA.57.737.1998, PhysRevLett.85.1762.2000, preskill2018quantum}]).

	
\begin{figure*}[tbp]
    \includegraphics[width=\linewidth]{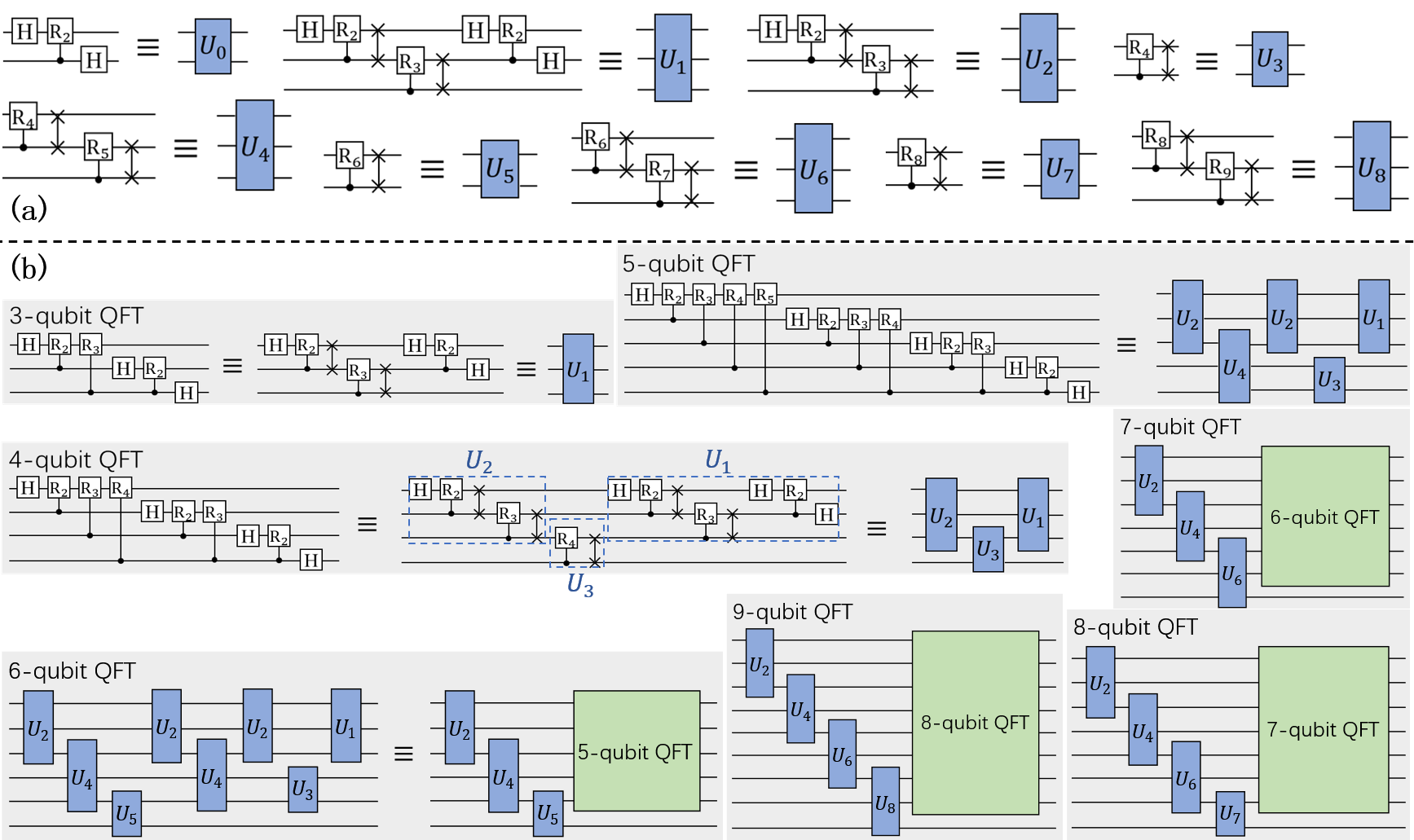}
    \caption{(Color online) (a) The nine elementary gates \R{$\{U_m\}$ ($m=0, \ldots, 8$)} in the $3$-qubit \R{QuVIS} for compiling the $N$-qubit QFT circuits for $N \leq 9$. \R{We use ``H'' to denote Hadamard gate, use two crosses connected by a vertical line to denote swap gate, and use ``$R_{p}$'' connected to a dot to denote controlled phase shift gate with the phase $\theta=\pi / 2^{p}$. The error and the time used to implement each elementary gate are shown in Table~\ref{U-tabel}. By adding necessary swap gates, the $N$-qubit QFT circuits compiled by the $3$-qubit QuVIS are shown in (b).}}
    \label{fig-VIS}
\end{figure*}

\begin{figure}[tbp]
	\centering
	\includegraphics[angle=0,width=0.9\linewidth]{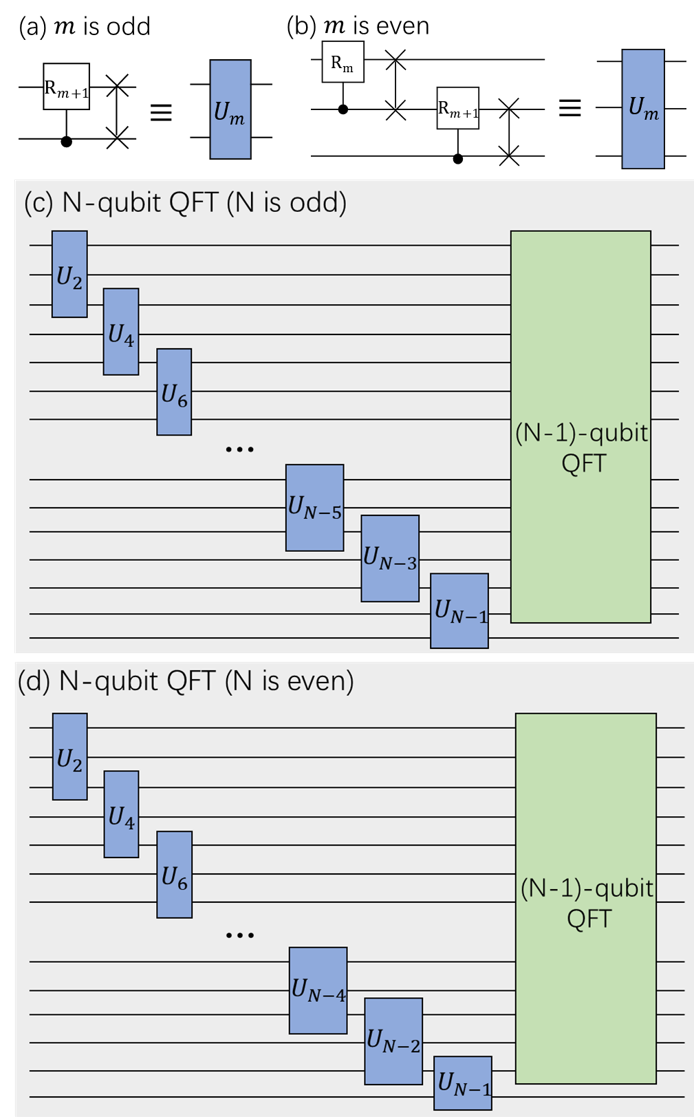}
	\caption{\RR{(Color online) The elementary gates for compiling the QFT circuits with 3-qubit QuVIS can be recursively derived.  Panels (a) and (b) show the definitions of the $m$-th elementary gate $U_{m}$ for $m \geq 3$ with odd and even $m$, respectively. The circuit after compiling the $N$-qubit QFT circuit can also be derived as illustrated in Panel (c) for odd $N$ and (d) for even $N$.}}
	\label{fig-UT}
\end{figure}

Aiming at higher efficiency and accuracy, we here propose the \R{quantum variational instruction set (QuVIS)} for compiling the quantum circuits. A \R{QuVIS} is defined as the flexibly-designed multi-qubit quantum gates that can be realized by controlling the magnetic pulses imposed on the interacting spins in a quantum hardware. The pulse sequences are variationally determined using the fine-grained time optimization (FGTO) algorithm~\cite{lu2021preparation}, which manages to efficiently realize the given multi-qubit unitary transformations. \RR{We take QuMIS~\cite{fu2017experimental} as an example to compare the performances (error and time cost). Our results show that QuVIS} significantly reduces the number of the elementary gates in the compiled circuit, thus suppresses the accumulation of errors and the time cost. These advantages of \R{QuVIS} are demonstrated on compiling the circuits of $N$-qubit quantum Fourier transformation (QFT)~\cite{ekert1996quantum, jozsa1998quantum, weinstein2001implementation} and multi-qubit swap circuits. We show the elementary gates of the \R{QuVIS} designed for the $N$-qubit QFT in Fig.~\ref{fig-VIS} (a) and the compiled circuits for $N=3, \ldots, 9$ in (b). Thanks to the generality and stability of FGTO on realizing unitary transformations, \R{QuVIS} can be adaptively defined for different quantum hardware with various interaction types (e.g., Ising or Heisenberg interactions), connectivities, and strengths among the qubits, \RR{according to the considered quantum hardware.}

\section{Variational instruction set}

To realize a target \RR{unitary transformation} $\hat{U}$, we optimize the adjustable parameters in the time-dependent Hamiltonian $\hat{H}(t)$ so that the time evolution operator in the duration $T$ optimally gives $\hat{U}$, i.e., $\hat{U} \simeq e^{-i\int_{0}^{T} \hat{H}(t) dt}$. We take the Plank constant $\hbar=1$ for simplicity. In many existing quantum hardware, the adjustable parameters of the Hamiltonian concerns the one-body terms, i.e., \RR{the magnetic pulses~\cite{vandersypen2001experimental, RevModPhys.76.1037, niu_universal_2019, PRXQuantum.2.040324}. We here take the Ising model with time-dependent transverse fields for demonstration.} The Ising Hamiltonian can be written as
\begin{equation}
\hat{H}(t) = \sum_{nn'} J_{nn'} \hat{S}^z_{n} \hat{S}^z_{n'} - 2\pi\sum_n [ h^x_n(t) \hat{S}^x_n + h^y_n(t) \hat{S}^y_n],
\label{eq-H}
\end{equation}
with $\hat{S}^{\alpha}_n$ the spin operator in the $\alpha$ direction ($\alpha=x, y, z$), $J_{nn'}$ the coupling constants between the $n$-th and $n'$-th spins, and $h^{\alpha}_n(t)$ the adjustable magnetic pulses along the spin-$\alpha$ direction on the $n$-th spin at the time $t$. The goal becomes optimizing $h^{\alpha}_n(t)$ to minimize the difference
\begin{equation}
\varepsilon = \left|\hat{U} - e^{-i\int_{0}^{T} \hat{H}(t) dt} \right|
\label{eq-st}
\end{equation}
\RR{where $\left| \ast \right|$ is the Frobenius norm.} Such optimizations can be efficiently implemented by the gradient-descent methods even when $\hat{U}$ concerns multiple qubits. 


\RR{Due to the generality of the optimization scheme, we are able to consider additional restrictions in the optimization process. For instance, we may restrict that only the magnetic field in either the x or y direction can be imposed at each time, or the strength of magnetic fields should be limited to a certain range. Such restrictions (and many restrictions in the realistic hardware) will not break the automagical differentiation chain, and thus can be readily considered in the optimization.}

We utilize the fine-grained time optimization (FGTO)~\cite{lu2021preparation} to optimize the pulse sequences for the target gates. The idea is to avoid being trapped in local minima by gradually fine-graining the time discretization. \R{The validity of this strategy has been demonstrated on the state-preparation tasks.} We take the Trotter-Suzuki form~\cite{trotter1959product, suzuki1976generalized} and discretize the total time $T$ to $\tilde{K}$ identical slices. The evolution operator can be approximated as
\begin{eqnarray}
\hat{U}(T) = e^{-i\tau \hat{H} (\tilde{K}\tilde{\tau})} \ldots  e^{-i\tilde{\tau} \hat{H} (2\tilde{\tau})} e^{-i\tilde{\tau} \hat{H} (\tilde{\tau})},
\label{eq-psit1}
\end{eqnarray}
with $\tilde{\tau} = \frac{T}{\tilde{K}}$ that controls the Trotter-Suzuki error. For varying the magnetic fields, we introduce $\tau = \kappa \tilde{\tau}$ with $\kappa$ a positive integer, and assume $h^{\alpha}_n(t)$ to take the constant value $h^{\alpha}_n(t) = h^{\alpha}_{n,k}$ during the time of $(k-1)\tau \leq t < k\tau$ (with $k=1, \cdots, K$ and $K=\frac{T}{\tau}$). In other words, $\tau$ controls the maximal frequency of the magnetic pulses, and the magnetic fields are allowed to change for $K$ times in the controlling duration. \R{During the optimization, $\tau$ is reduced gradually to increase the fineness of time discretization.We start from a relatively large $\tau$ and reduce it to ${\tau/2}$ when $\{h^{\alpha}_{n,k}\}$ converge. The length (i.e., the dimension of the index $k$) of the pulse sequences will be doubled. At the beginning of the optimization with a new (smaller) $\tau$, the pulse sequences are initialized as $h^{\alpha}_{n,2k'-1} = h^{\alpha}_{n,2k'} \leftarrow h^{\alpha}_{n,k'}$.} 

The magnetic fields are updated as
\begin{eqnarray}
h^{\alpha}_{n,k} \leftarrow h^{\alpha}_{n,k} - \eta \frac{\partial \varepsilon}{\partial h^{\alpha}_{n,k}},
\label{eq-updatehd}
\end{eqnarray}
where the gradients $\frac{\partial \varepsilon}{\partial h^{\alpha}_{n,k}}$ can be obtained by, e.g., the automatic differentiation technique in Pytorch~\cite{PyTorch}. We use the optimizer Adam~\cite{KB15Adam} to dynamically control the learning rate $\eta$.

The \R{QuVIS} for different quantum circuits can be defined flexibly. Specifically, we call a \R{QuVIS} to be \RR{$\tilde{N}$}-qubit when the elementary gates therein are at most \RR{$\tilde{N}$}-qubit. Let us take the QFT as an example, which belongs to the most frequently used circuits in implementing quantum algorithms including Shor~\cite{365700} and Grover algorithms ~\cite{PhysRevLett.79.325}. Fig.~\ref{fig-VIS} (a) gives the $3$-qubit \R{QuVIS} for the $N$-qubit QFT with $N \leq 9$, and (b) shows the circuits after compiling. The magnetic fields to realize each elementary gate is obtained by the algorithm explained above. 

\RR{The complexity of obtaining the magnetic fields on a classical computer (i.e., optimization complexity) increases exponentially with $\tilde{N}$ (the maximal number of qubits in the elementary gates of QuVIS). This optimization complexity is independent on the number of qubits $N$ in the circuit that is to be compiled. In comparison, we may consider the whole quantum circuit as a large unitary transformation, and use FGTO to minimize the distance between this unitary transformation and the time-evolution operator. We dub such a simple and ``brute-force'' scheme as direct control, which will be used later as a baseline. In this case, we need to simulate the time evolution of an $N$-qubit system, thus the optimization complexity of the direct control scheme increases exponentially with $N$}. 

Here we focus on the \R{QuVIS} with \RR{$\tilde{N}=2$} and $3$, which already exhibits significant advantages on efficiency and accuracy (see the benchmark results). Be aware that one can use a desktop computer to access the \R{QuVIS's} for \RR{$\tilde{N} \leq 6$} without any problems. Other than QFT, \R{QuVIS} can also be designed flexibly for different quantum circuits or algorithms. One may find more details on the optimization and the controlling sequences for realizing the elementary gates of QuVIS in the Supplemental Material~\cite{SM}.

\RR{The elementary gates or most of them in a QuVIS can be derived recursively. Taking the QuVIS for QFT as an example, the elementary gates from $U_0$ to $U_2$ are designed manually. Clear regularity appears to derive the rest of gates recursively. For the definitions of the elementary gates in QuVIS, the $m$-th gate $U_{m}$ is composed of the rotational gate $R_{m+1}$ and a SWAP gate when $m$ is odd. For an even $m$, $U_{m}$ is composed of two rotational gates ($R_{m+1}$ and $R_{m}$) and two SWAP gates, see Fig.~\ref{fig-UT}(a) and (b). Considering to compile the $N$-qubit QFT by QuVIS, the circuit consists of $U_2$, $U_4$, $U_6$, ..., $U_{N-5}$, $U_{N-3}$, $U_{N-1}$, and the ($N$-1)-qubit QFT circuit, when $N$ is odd [Fig.~\ref{fig-UT}(c)]. When $N$ is even, the $N$-qubit QFT circuit consists of $U_2$, $U_4$, $U_6$, ..., $U_{N-4}$, $U_{N-2}$, $U_{N-1}$, and the ($N$-1)-qubit QFT circuit [Fig.~\ref{fig-UT}(d)].}

\section{Benchmark results}

Below, we take the Hamiltonian for time evolution to be the nearest-neighbor Ising chain, where the coupling constants satisfy
\begin{eqnarray}
  J_{nn'}=
  \left\{
  	\begin{array}{rcl}
		2\pi & \text{ for }n' = n+1\\
		0 &\text{otherwise}
	\end{array}. \right.
\end{eqnarray}
\RR{In our demonstration,} we fix the magnetic fields along the spin-z direction as zero, and allow to \RR{independently} adjust the fields along the spin-x and y directions. Such a case often appears in the controlling by the radio-frequency pulses~\cite{RevModPhys.76.1037, PhysRevX.7.031011}.

Table~\ref{U-tabel} shows the time cost $T$ for realizing the elementary gates $\{\hat{U}_m\}$ ($m=0, \ldots, 8$) by FGTO (second row). For comparison, we also estimate the time cost by compiling each gate to the product of the elementary gates in QuMIS (third row). 
\RR{To conveniently and fairly compare the time cost, we take the $T$ when the error [Eq.~(\ref{eq-st})] decreases to about $O({10}^{-2})$.} The time cost of implementing the elementary gate in the \R{QuVIS} using FGTO is significantly shorter than that by compiling them to the product of the elementary gates in QuMIS. \RR{Note that in general, the loss function will decrease as the total time duration $T$ increases, until the limit of the optimization scheme is reached. Such a limit is determined by many factors including the gradient step (learning rate) and other optimization tricks (such as the optimizer, for which we choose Adaptive Moment Estimation~\cite{kingma2017adam}). In our simulations, the loss function will eventually converges to about $O({10}^{-6})$ (see Fig. S2 in the supplemental material.} 

\begin{table}[tbp]
    \setlength{\tabcolsep}{3pt}
	\centering
		\caption{The time cost $T$ to implement the elementary gates $\{\hat{U}_{m}\}$ ($m=0, \ldots, 8$) of the $3$-qubit \R{QuVIS} [Fig.~\ref{fig-VIS}(a)] for QFT. The second row shows the results by directly taking $\{\hat{U}_{m}\}$ as the target gates in Eq.~(\ref{eq-st}), and the third row shows those by compiling $\{\hat{U}_{m}\}$ to the product of the elementary gates in QuMIS.}
		\label{U-tabel}
		\begin{tabular*}{8.7cm}{@{\extracolsep{\fill}}lcccccccccc}
			\hline\hline
			 $T$& $\hat{U}_0$ & $\hat{U}_1$ & $\hat{U}_2$ & $\hat{U}_3$ & $\hat{U}_4$ & $\hat{U}_5$ & $\hat{U}_6$ & $\hat{U}_7$ & $\hat{U}_8$ \\ \hline
			$\text{\R{QuVIS}}$ & 0.3 & 2.1 & 2.1 & 1.4 & 2.4 & 1.5 & 2.4 & 1.5 & 2.4 \\ \hline
			$\text{QuMIS}$ & 2.3 & 8.4 & 6.0 & 2.6 & 5.1 & 2.5 & 5.0 & 2.5 & 5.0 \\ \hline
			\hline
		\end{tabular*}
\end{table}

\begin{figure}[tbp]
	\centering
	\includegraphics[angle=0,width=0.9\linewidth]{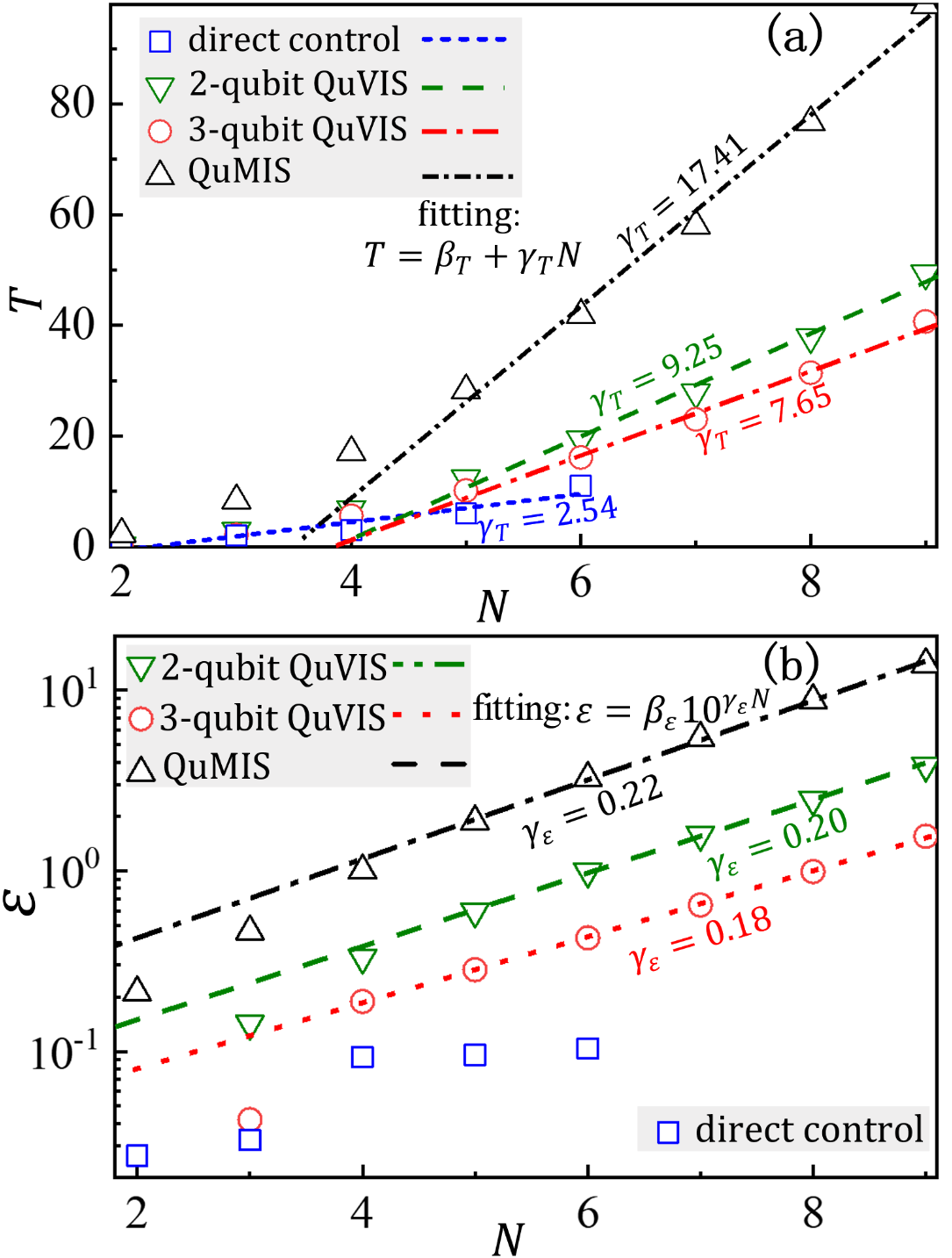}
	\caption{(Color online) (a) The time cost $T$ and (b) error $\varepsilon$ in realizing the $N$-qubit QFT using the direct control, 2-qubit \R{QuVIS}, 3-qubit \R{QuVIS}, and QuMIS. The dash lines give the linear fitting of the time cost $T$ versus $N$ [Eq.~(\ref{eq-fittingT})]. In the optimization for the direct control, the time cost is estimated under the condition that $\varepsilon$ is no more than $10^{-1}$. For the 2-qubit \R{QuVIS}, 3-qubit \R{QuVIS}, and QuMIS, $\varepsilon$ changes exponentially with $N$ [Eq.~(\ref{eq-fittingE})]. \RR{Note the fittings are performed using the data for $N \geq 5$.}}
	\label{fig-data}
\end{figure}

In Fig.~\ref{fig-data}, we demonstrate the time costs $T$ and the error $\varepsilon$ in realizing the circuits for $N$-qubit QFT. \RR{The direct control scheme is used as a baseline method to compare with QuVIS and QuMIS. Though it exhibits the lowest error, its disadvantage is that the computational cost increases exponentially with the number of qubits $N$ in the quantum circuit to be compiled. Thus, it is not feasible to apply to the circuits of large sizes.} 

Compared with QuMIS, significant reductions on both the time cost $T$ and error $\varepsilon$ are demonstrated by using the $2$- and $3$-qubit \R{QuVIS} for compiling. Since $T$ is determined by the number of elementary gates and the time to realize each of them, it is approximately linear to the number of qubits $N$. We have
\begin{eqnarray}
	T = \gamma_{T} N + \beta_{T},
	\label{eq-fittingT}
\end{eqnarray}
with the slope $\gamma_{T} \simeq 17.41$, $9.25$, and $7.65$ for QuMIS and \RR{$\tilde{N}$}-qubit \R{QuVIS} with \RRR{$\tilde{N} = 2$ and $3$}, respectively.

Since each elementary gate inevitably introduces certain error (fixed to be $O(10^{-2})$ in our simulations), the error $\varepsilon$ for the whole circuit generally accumulates exponentially as $N$ increases. We have
\begin{eqnarray}
	\varepsilon =  \beta_{\varepsilon}e^{\gamma_{\varepsilon} N},
	\label{eq-fittingE}
\end{eqnarray}
with the exponent coefficient $\gamma_{\varepsilon}=0.22$, $0.2$, and $0.18$ for QuMIS and \R{$\tilde{N}$}-qubit \R{QuVIS}. A reduction of $\gamma_{\varepsilon}$ indicates an algebraic improvement, essentially because that the number of elementary gates (i.e., depth) of the compiled circuit is reduced by increasing \RR{$\tilde{N}$}. 

The key advantage of \R{QuVIS} is from the ``end-to-end'' optimization strategy for the magnetic pulses. When a \RR{unitary transformation} is compiled to the product of several gates, the conventional schemes require accurate implementations of all gates. However, we actually care about the \RR{unitary transformation} itself but not any intermediate results within the compiled circuit.

An \RR{$\tilde{N}$}-qubit \R{QuVIS} is designed by dividing the target circuit into many sub-circuits [Fig.~\ref{fig-VIS}(b)], where each sub-circuit is at most \RR{$\tilde{N}$}-qubit and the total number of the sub-circuits should be as small as possible. These sub-circuits define the elementary gates in the \R{QuVIS} [Fig.~\ref{fig-VIS}(a)]. The magnetic pulses are optimized by directly finding the optimal path to each elementary gate, without considering the intermediate results within the corresponding sub-circuit. Meanwhile, a properly designed \R{QuVIS} will significantly reduce the number of elementary gates in a compiled circuit. For these reasons, the circuit compiled by a \R{QuVIS} exhibits much less error and time cost compared with that by a standard QIS.

To provide an explicit demonstration, we show in Fig.~\ref{fig-1} the error $\varepsilon(t)$ in the controlling duration
\begin{eqnarray}
	\varepsilon(t) = \left|\hat{U}(\theta) - e^{-i\int_{0}^{t} \hat{H}(t') dt'} \right|.
	\label{eq-errorT}
\end{eqnarray}
\RR{This quantity gives the distance at a certain time point $t$ with $0\le t\le T$, where the magnetic fields are still optimized by minimizing $\varepsilon(T)$ [Eq.~(\ref{eq-st})].} As an example, we take $\hat{U}(\theta)$ to be the controlled phase shift gate
\begin{eqnarray}
	 \hat{U}(\theta) =
  \left[
  \begin{matrix}
   1 & 0 & 0 & 0 \\
   0 & 1 & 0 & 0 \\
   0 & 0 & 1 & 0 \\
   0 & 0 & 0 & e^{i\theta}
  \end{matrix}
  \right].
\label{eq-CU}
\end{eqnarray}

\begin{figure}[tbp]
	\centering
	\includegraphics[angle=0,width=0.95\linewidth]{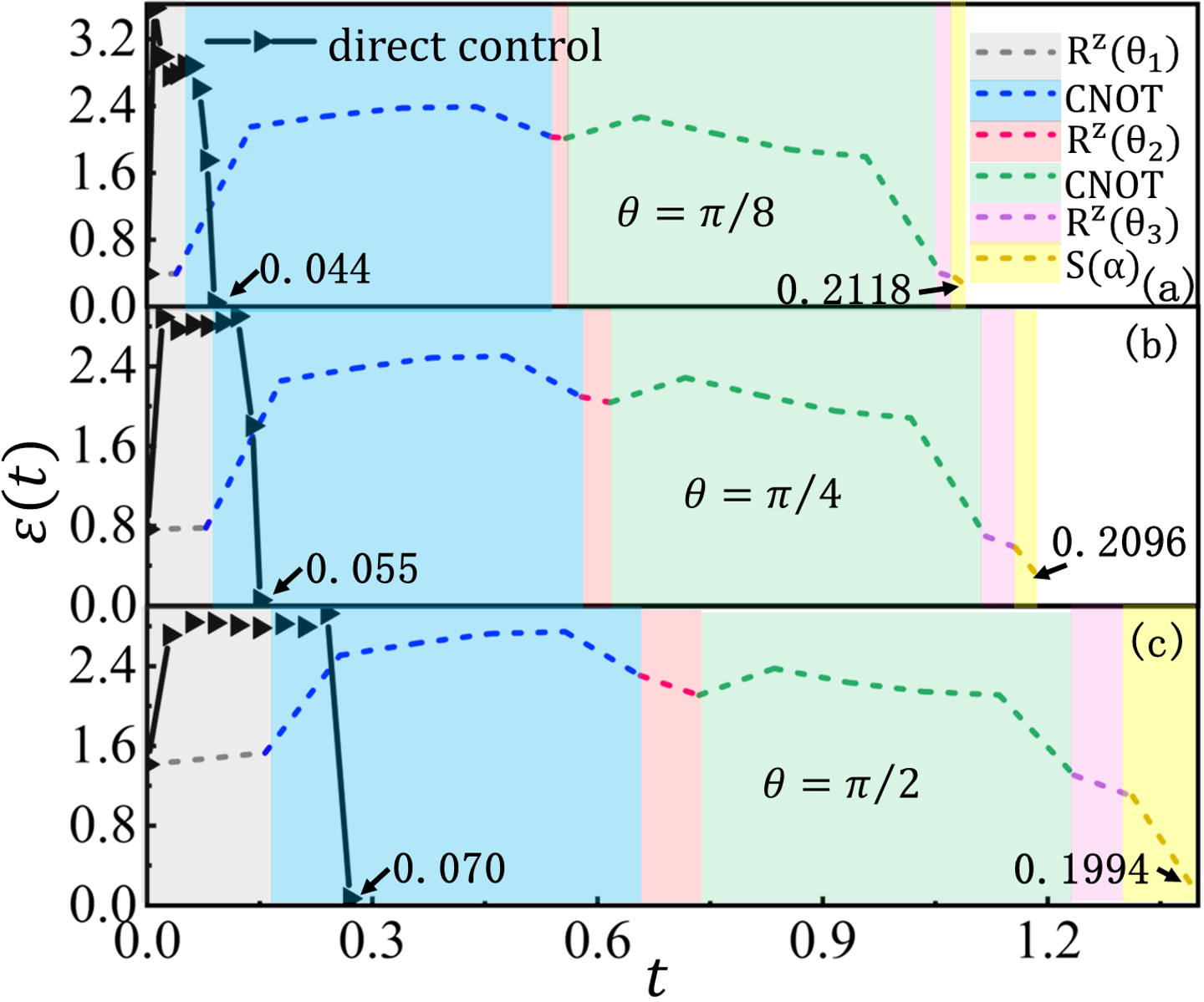}
	\caption{(Color online) The error $\varepsilon(t)$ [Eq.~(\ref{eq-errorT})] versus the evolution time $t$ to realize the controlled phase shift gate [Eq.~(\ref{eq-CU})]. The dashed lines and the solid lines with triangles show the $\varepsilon(t)$ by the QuMIS and direct control, respectively. The colored shadows indicate the time cost for realizing the gates in the right-hand-side of Eq.~(\ref{eq-R}).}
	\label{fig-1}
\end{figure}

For the phase shift $\theta=\frac{\pi}{8}, \frac{\pi}{4}$, and $\frac{\pi}{2}$, Fig. \ref{fig-1} compares the error $\varepsilon(t)$ [Eq.~(\ref{eq-errorT})] by directly minimizing the distance to the target gate (direct control) and by the standard compiling. Using QuMIS, $\hat{U}(\theta)$ is decomposed to the product of the single-qubit rotation gates \R{$\hat{R}^{z}$} and CNOT $\hat{C}$, \R{which can be formally written as}
\begin{eqnarray}
	\hat{U}(\theta) = \hat{S}(\alpha) \hat{R}^{z}(\theta_1)\hat{C}\hat{R}^{z}(\theta_2)\hat{C}\hat{R}^{z}(\theta_3),
	\label{eq-R}
\end{eqnarray}
\R{where $\hat{R}^{z}(\theta_1)$, $\hat{R}^{z}(\theta_2)$ and $\hat{R}^{z}(\theta_3)$ are single-qubit rotations along the spin-z direction satisfying $\hat{R}^{z}(\theta_1)\hat{R}^{z}(\theta_2)\hat{R}^{z}(\theta_3)=I$, and $\hat{S}(\alpha) = e^{i\alpha}$ a phase factor~\cite{barenco1995elementary}. Note that all single-qubit gates in Eq.~(\ref{eq-R}) are acted on the second qubit. In other words, when the control qubit (the first one here) is in the state $|1\rangle$, the target qubit (second one) will be acted by $\hat{S}(\alpha)\hat{R}^{z}(\theta_1)\hat{X}\hat{R}^{z}(\theta_2)\hat{X}\hat{R}^{z}(\theta_3)$ with $\hat{X}$ is Pauli-x operator.}

\begin{figure}[tbp]
	\centering
	\includegraphics[angle=0,width=0.9\linewidth]{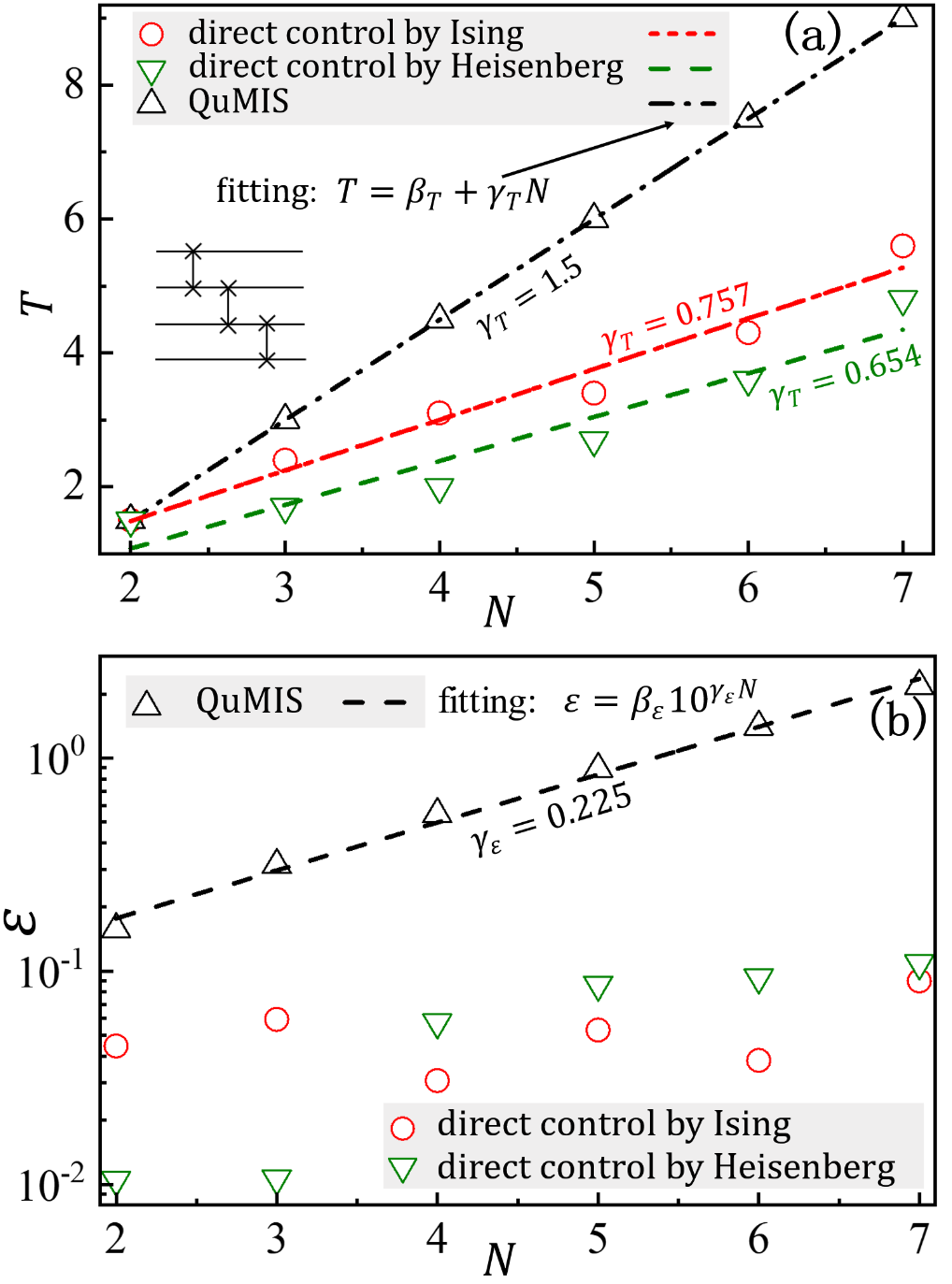}
	\caption{(Color online) (a) The time cost $T$ and (b) the corresponding error $\varepsilon$ for the $N$-qubit swap by direct control with Ising and Heisenberg interactions. The inset in (a) illustrates the circuit of $4$-qubit swap. The data of QuMIS with Ising interactions are also given for comparison. The fitting functions of $T$ and $\varepsilon$ are given in Eqs.~(\ref{eq-fittingT}) and (\ref{eq-fittingE}), respectively.}
	\label{fig-2}
\end{figure}

The time costs of realizing the elementary gates in QuMIS are illustrated by the colored shadows. The time cost of direct control is indicated by the x-coordinate of the last triangle, which is about five times shorter than QuMIS. Note for a single-qubit rotation \R{$\hat{R}^{\alpha}(\theta)$}, it can be written as the one-body evolution operator with the magnetic field along the corresponding direction, i.e., \R{$\hat{R}^{\alpha}(\theta) = e^{-i\theta \hat{S}^{\alpha}} \Leftrightarrow \hat{U}(h^{\alpha}, T) = e^{-i T h^{\alpha} \hat{S}^{\alpha}}$}. Therefore, the time cost of \R{$\hat{R}^{\alpha}(\theta)$} is estimated as $T= \frac{\theta}{h^{\alpha}}$. Without losing generality, we here take $h^{\alpha}=10$ to estimate the time costs of single-qubit rotations.

An important observation is that even the time cost of a single CNOT ($T=0.5$ theoretically given in Refs.~\cite{li2013time, sun2020time}) is longer than that of $\hat{U}(\theta)$ by direct control. Meanwhile, direct control also exhibits much lower errors with $\varepsilon \sim O(10^{-2})$. For QuMIS, the error accumulates and finally reaches $O(10^{-1})$ that is about ten times larger than that by direct control. Therefore, from the perspective of \R{QuVIS}, it becomes less efficient and accurate by decomposing the $\hat{U}(\theta)$ into the product of CNOT and the single-qubit rotations.

The pulse sequences can be optimized for the quantum platforms with different interactions. Fig.~\ref{fig-2} shows the time cost $T$ and the corresponding error $\varepsilon$ for the $N$-qubit swap circuit using direct control with Ising and Heisenberg interactions. The circuit swaps the first qubit to the last [see the inset of Fig.~\ref{fig-2}(a)]. The time $T$ is estimated by keeping the error of each elementary gate to be $O(10^{-1})$ or less. Linear scaling of $T$ given by Eq.~(\ref{eq-fittingT}) is observed for both kinds of interactions. Thanks to the flexibility of the optimization algorithm, the pulse sequences can be obtained for any types and strengths of the interactions, and the error of realizing the elementary gates can be readily estimated.

\section{Summary}

We here propose the \R{quantum variational instruction set (QuVIS)} for the efficient quantum computing based on the dynamics of the interacting spin systems controlled by pulse sequences of magnetic fields. The key idea of \R{QuVIS} is by flexibly defining the multi-qubit elementary gates, where we ignore the intermediate processes but optimize the magnetic fields to directly realize the target unitary transformations. By taking the $N$-qubit quantum Fourier transformation as an example, significant reductions on the time cost and error accumulations are demonstrated compared with the standard quantum instruction set. \R{QuVIS} provides a flexible quantum compiling scheme generally for the quantum platforms with known interactions. For the cases where the interactions are unknown, one can combine with the methods that estimate the interactions using, e.g., the machine learning of the local observables and reduced density matrices~\cite{xin_local-measurement-based_2019, Li_2020, MTR21MLHamilt}.

%




\section*{Acknowledgment} This work is supported by NSFC (Grant No. 12004266, No. 11834014, No. 12075159, and No. 12171044), Beijing Natural Science Foundation (Grant No. Z190005), Foundation of Beijing Education Committees (Grant No. KM202010028013), the key research project of Academy for Multidisciplinary Studies, Capital Normal University, and the Academician Innovation Platform of Hainan Province.

\bibliography{name}

\clearpage
\onecolumngrid
\appendix

\section*{Supplemental material of ``Quantum compiling with variational instruction set for accurate and fast quantum computing"}

In this supplemental material, we provide more information on realizing $N$-qubit quantum Fourier transformation (QFT) by direct control, and the sequences of magnetic fields for realizing the elementary gates $\{\hat{U}_{m}\}$ ($m=0, \ldots, 8$) in the $3$-qubit quantum variational construction set (QuVIS) for QFT.

As an example, we take the time-dependent Hamiltonian as the \R{quantum Ising model with transverse fields}, which reads
\begin{eqnarray}
\hat{H}(t) = \sum_{n=1}^{N-1} 2\pi \hat{S}^z_{n} \hat{S}^z_{n+1} - \sum_{n=1}^{N} [2\pi h^x_n(t) \hat{S}^x_n + 2\pi h^y_n(t) \hat{S}^y_n].
\label{eq-H}
\end{eqnarray}
Its adjustable parameters only concern the one-body terms, i.e., the magnetic fields along the spin-x and y directions. The evolution operator for the time duration $T$ is approximated by Trotter-Suzuki decomposition (we take Plank constant $\hbar=1$ as the energy scale) as
\begin{eqnarray}
e^{-i\int_{0}^{T} \hat{H}(t) dt} \simeq \prod_{k=1}^K \prod_{\tilde{k}=1}^{\lfloor \tilde{\tau} / \tau \rfloor} \exp{\left[-i 2\pi \tau \left(\sum_{n=1}^{N-1} \hat{S}^z_{n} \hat{S}^z_{n+1} + \sum_{n=1}^{N} h^x_{n,k} \hat{S}^x_n + \sum_{n=1}^{N} h^y_{n,k} \hat{S}^y_n\right) \right]},
\label{eq-U}
\end{eqnarray}
with $\tau=T/K$ controlling the maximal frequency of the pulse sequences, $\tilde{\tau} \leq \tau$ the Trotter step that controls the Trotter error, and $\lfloor \ast \rfloor$ the round-down operation. Specifically speaking, we assume that the magnetic field on the $n$-th qubit along the $\alpha$ direction for $(k-1)\tau<t \leq k\tau$ takes the constant value $h^{\alpha}_{n,k}$. An extra term $\exp{\left[-i 2\pi \tau' \hat{H}(t) \right]}$ will be added for each $k$ if $\tilde{\tau} / \tau$ is not an integer, with $\tau'=\tau - \tilde{\tau} \lfloor \tilde{\tau} / \tau \rfloor$. To realize a target \R{unitary transformation} $\hat{U}$ (which can be a circuit or an elementary gate in a QuVIS), the goal is to optimize $\{h^{\alpha}_{n,k}\}$ so that the evolution operator approximately gives the \R{unitary transformation}, i.e., $\hat{U} \simeq e^{-i\int_{0}^{T} \hat{H}(t) dt}$. The error can be characterized by the distance between these two operators as 
\begin{eqnarray}
\varepsilon &=& \left|\hat{U} - e^{-i\int_{0}^{T} \hat{H}(t) dt} \right|.
\label{eq-loss}
\end{eqnarray}

The minimization of $\varepsilon $ suffers from the local minima. In the reference [Phys. Rev. A 104, 052413 (2021)],  the fine-grained time optimization (FGTO) is proposed. Its key idea is to gradually decrease $\tau$ the discretization of time. We start from a relatively large $\tau$ and reduce it to $\tau/2$ when $\{h^{\alpha}_{n,k}\}$ converge. The length (i.e., the dimension of the index $k$) of the pulse sequences will be doubled. At the beginning of the optimization with a new (smaller) $\tau$, the pulse sequences are initialized as $h^{\alpha}_{n,2k'-1} = h^{\alpha}_{n,2k'} \leftarrow h^{\alpha}_{n,k'}$.

\begin{figure}[tbp]
    \centering
    \includegraphics[width=7cm]{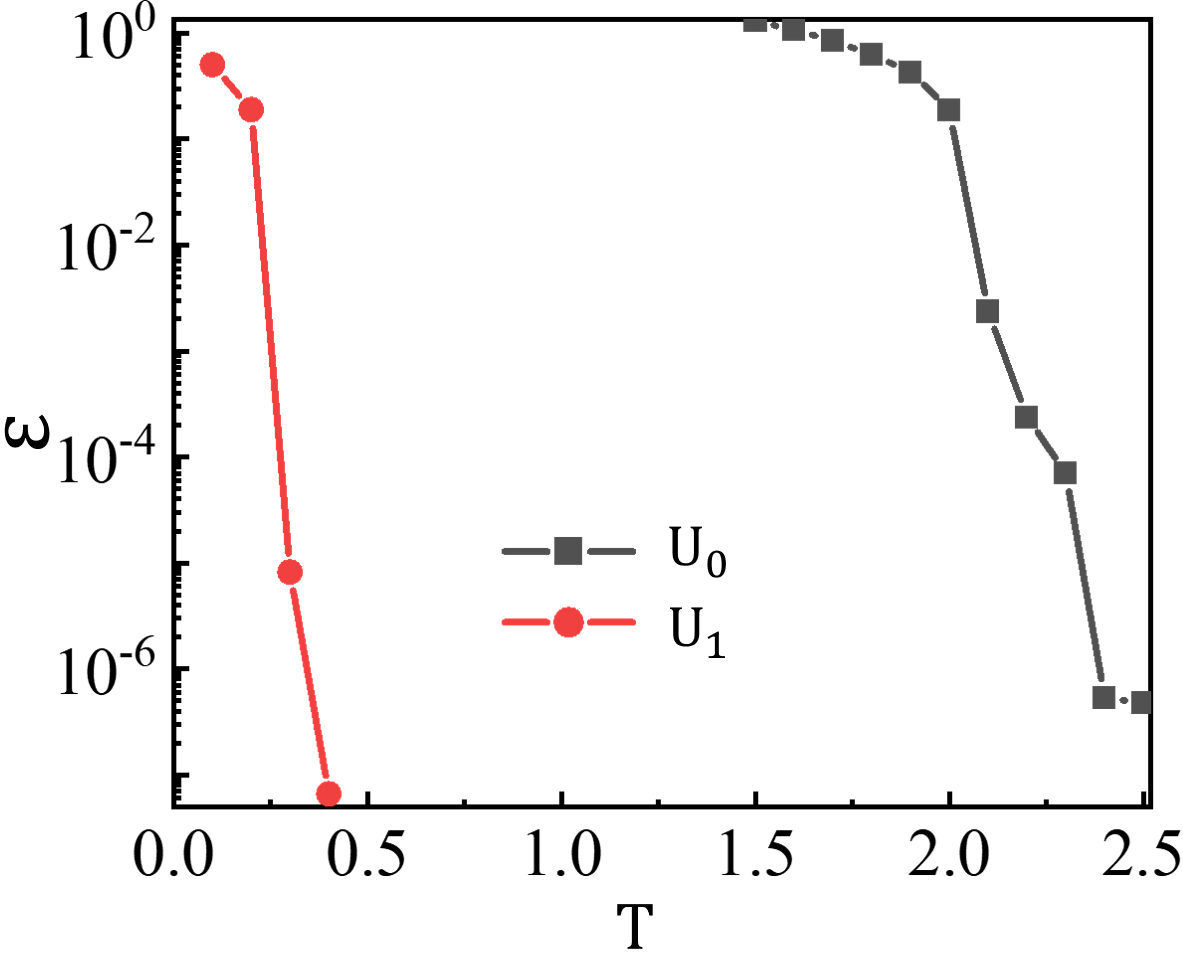}
    \caption{\R{(Color online) The error $\varepsilon$ [Eq.~(\ref{eq-loss})] against $T$ for the elementary gate $U_0$ and $U_1$ of 3-qubit QuVIS.}}
    \label{fig-3}
\end{figure}

\R{Fig.~\ref{fig-3} shows the error $\varepsilon$ against $T$ for the elementary gate $U_0$ and $U_1$ of the 3-qubit QuVIS, where $\varepsilon$ gradually decreases with $T$ and finally converges to about $O(10^{-6})$. Such a convergence is relevant to the gradient step, the optimization tricks, and etc.}

\begin{figure}[hbp]
    \centering
    \includegraphics[width=7cm]{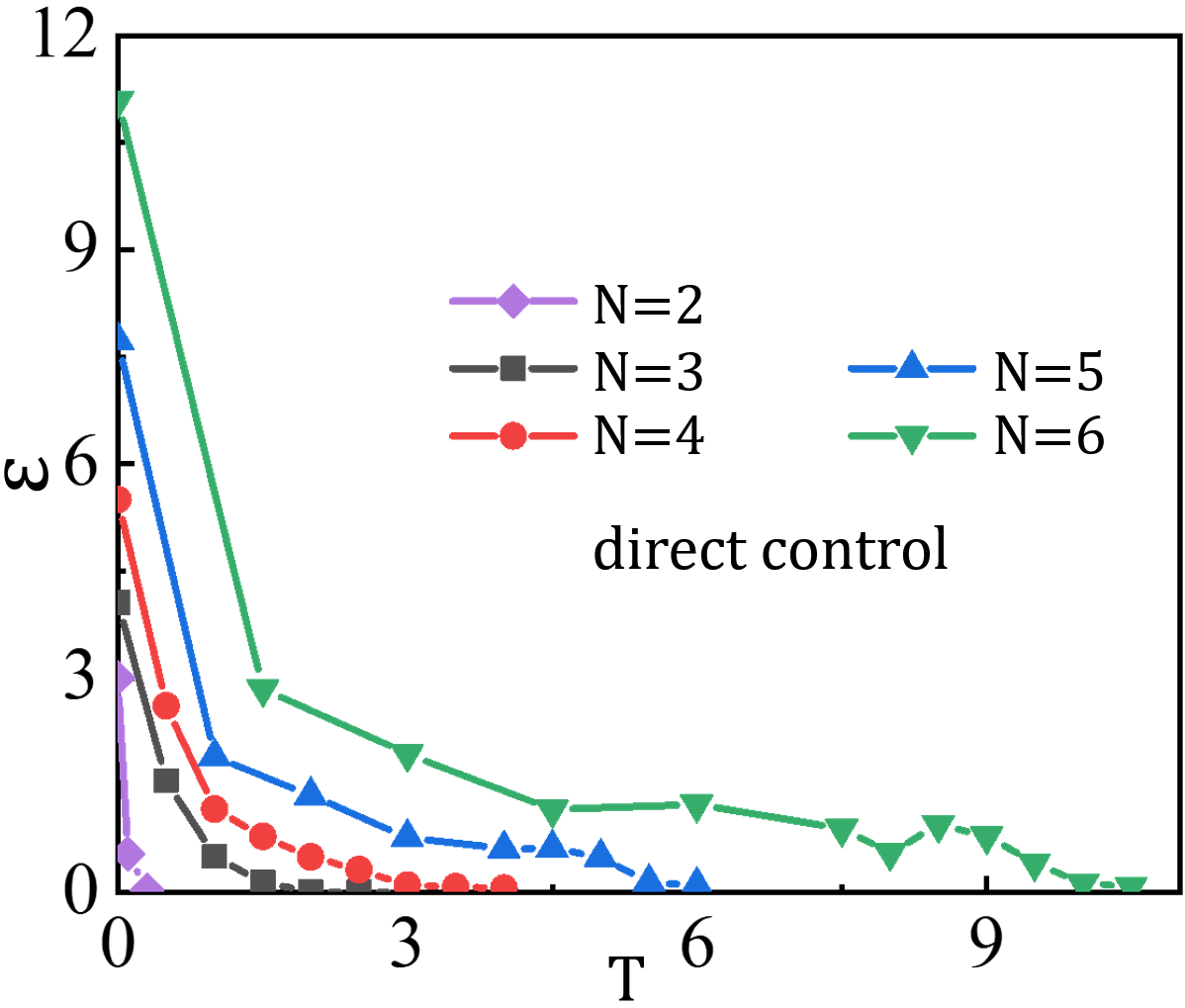}
    \caption{(Color online) The error $\varepsilon$ [Eq.~(\ref{eq-loss})] with different total evolution duration $T$ for the $N$-qubit QFT with $N=2, \cdots, 6$ by the direct control. In general, one can get lower $\varepsilon$ by increasing $T$. Longer evolution time is required to reach a preset error if $N$ increases. The errors from the direct control are controlled to be about $10^{-2}$.}
    \label{fig-1}
\end{figure}


\begin{figure}[tbp]
    \centering
    \includegraphics[width=18cm]{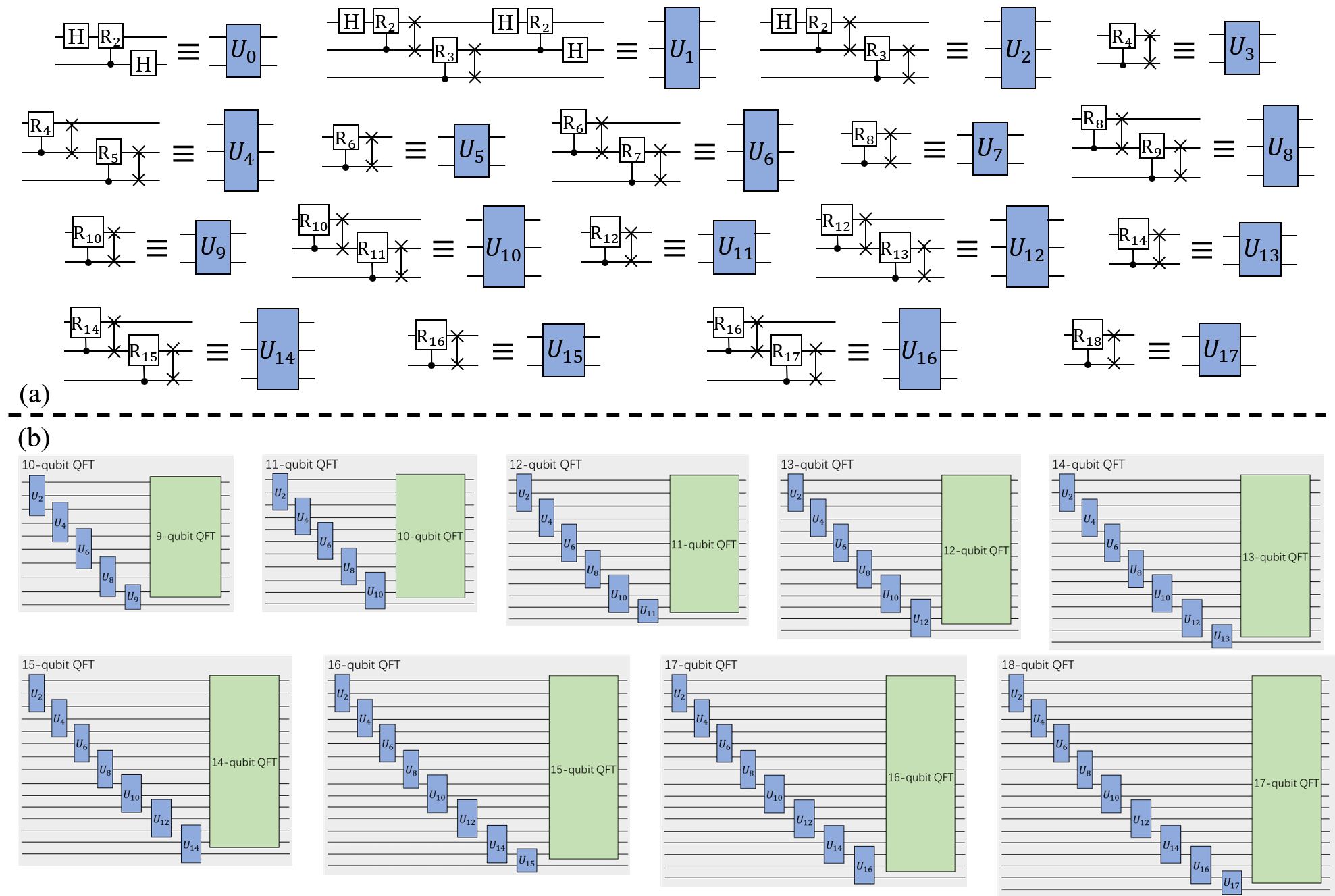}
    \caption{\R{(Color online) (a) The 18 elementary gates in the $3$-qubit QuVIS for compiling the $N$-qubit QFT circuits. (b) The quantum circuits obtained by compiling the $N$-qubit QFT using $3$-qubit QuVIS.}}
    \label{fig-4}
\end{figure}

    \begin{table}[ht]
	\centering
		\caption{The magnetic field $\{h^x_{n,k}\}$ and $\{h^y_{n,k}\}$ to realize the $U_0$. We take the total time $T=0.3$, $K=30$, and thus the Trotter step $\tau=0.3/30$.} \label{table-1}
\begin{tabular*}{13cm}{@{\extracolsep{\fill}}lcccccc}
            \\
			\hline\hline
			\textbf{$U_0$} & $n=1(x)$ & $n=2(x)$ & $n=1(y)$ & $n=2(y)$  \\ \hline
            k = 1 & -6.3436 & -3.7902 & -10.6531 & -2.2952 \\ \hline
            k = 2 & -3.6063 & -2.5666 & -7.8483 & -0.9786 \\ \hline
            k = 3 & -0.8042 & -1.4897 & -2.7742 & -0.5759 \\ \hline
            k = 4 & 1.4108 & 0.2407 & -2.0954 & 1.4366 \\ \hline
            k = 5 & -0.2386 & 0.8113 & -0.7169 & -0.8223 \\ \hline
            k = 6 & -0.6567 & -0.4742 & -1.0502 & -0.1966 \\
\end{tabular*}
	\end{table}
	
	\begin{table}[ht]
	\centering
		\caption{The magnetic field $\{h^x_{n,k}\}$ and $\{h^y_{n,k}\}$ to realize the $U_1$. We take the total time $T=2.1$, $K=210$, and thus the Trotter step $\tau=2.1/210$.} \label{tab-tabel}
		\scalebox{0.99}{
		\begin{tabular*}{13cm}{@{\extracolsep{\fill}}lcccccc}
            \\
			\hline\hline
			\textbf{$U_1$} & $n=1(x)$ & $n=2(x)$ & $n=2(x)$ & $n=1(y)$ & $n=2(y)$ & $n=2(y)$ \\ \hline
            k = 1 & -3.4147 & -3.8211 & -2.0099 & 16.7485 & -0.7455 & 3.0740 \\ \hline
            k = 2 & -0.8133 & 0.2982 & -0.9820 & 5.5036 & -0.4445 & 4.3913 \\ \hline
            k = 3 & 0.0319 & 0.3530 & -1.1379 & 2.4777 & 0.1856 & 4.3259 \\ \hline
            k = 4 & 1.0085 & 1.8844 & -0.8803 & -0.2038 & -0.5711 & 1.7134 \\ \hline
            k = 5 & 1.8824 & 2.4604 & -1.5054 & 2.3626 & 1.0999 & 1.3400 \\ \hline
            k = 6 & 0.0705 & 0.6606 & -1.6240 & -0.3138 & 1.3012 & 0.1534 \\
\end{tabular*}
		}
	\end{table}

	\begin{table}[tbp]
	\centering
	    \scalebox{0.99}{
		\begin{tabular*}{13cm}{@{\extracolsep{\fill}}lcccccccccccc}
            \\
			\hline\hline
			\textbf{$U_1$} & $n=1(x)$ & $n=2(x)$ & $n=3(x)$ & $n=1(y)$ & $n=2(y)$ & $n=3(y)$ \\ \hline
            k = 51 & -0.8192 & -0.7235 & -0.6687 & 0.7281 & 0.9322 & -0.7641 \\ \hline
            k = 52 & 0.2525 & -0.0066 & 1.0928 & -1.5057 & -1.9388 & -0.7399 \\ \hline
            k = 53 & -0.1577 & -1.1730 & -0.5157 & -0.2108 & 0.6388 & -0.1768 \\ \hline
            k = 54 & 0.0171 & 0.9999 & 0.0703 & -0.1239 & -0.6903 & 0.2969 \\ \hline
            k = 55 & -0.1934 & -0.8984 & -0.3649 & -0.1265 & 0.4941 & -0.8252 \\ \hline
            k = 56 & -0.5435 & 0.5608 & -0.0205 & 1.1567 & -2.0490 & 1.8442 \\
\end{tabular*}
		}
	\end{table}
	
	\begin{table}[tbp]
	\centering
	    \scalebox{0.99}{
		\begin{tabular*}{13cm}{@{\extracolsep{\fill}}lcccccccccccc}
            \\
			\hline\hline
			\textbf{$U_1$} & $n=1(x)$ & $n=2(x)$ & $n=3(x)$ & $n=1(y)$ & $n=2(y)$ & $n=3(y)$ \\ \hline
            k = 101 & -0.0430 & -0.1798 & -0.8335 & 0.2956 & 0.1394 & 0.1235 \\ \hline
            k = 102 & 0.1795 & 0.5883 & 1.3912 & -0.5283 & 0.2500 & -0.4987 \\ \hline
            k = 103 & 0.7849 & -0.6582 & -1.8932 & -0.8270 & 1.3872 & 0.1138 \\ \hline
            k = 104 & -0.2103 & -0.2454 & 0.6195 & 1.5247 & -0.8407 & -0.5403 \\ \hline
            k = 105 & -0.5493 & -0.0167 & -1.0297 & -0.8404 & 0.1436 & 0.3149 \\ \hline
            k = 106 & -0.0588 & 0.3188 & 0.3290 & 0.3744 & 0.2609 & 0.6131 \\
\end{tabular*}
		}
	\end{table}
	
	\begin{table}[tbp]
	\centering
	    \scalebox{0.99}{
		\begin{tabular*}{13cm}{@{\extracolsep{\fill}}lcccccccccccc}
            \\
			\hline\hline
			\textbf{$U_1$} & $n=1(x)$ & $n=2(x)$ & $n=3(x)$ & $n=1(y)$ & $n=2(y)$ & $n=3(y)$ \\ \hline
            k = 151 & -1.4397 & -0.8447 & -0.4617 & 0.9424 & 0.8580 & -1.4291 \\ \hline
            k = 152 & 0.4479 & -0.3353 & -0.0559 & -0.3454 & -0.3978 & -0.9241 \\ \hline
            k = 153 & -0.5748 & -0.6616 & 0.7613 & 0.7606 & -0.5344 & 0.6070 \\ \hline
            k = 154 & 0.5781 & -0.0446 & -0.3373 & -1.0824 & -0.2750 & -0.9404 \\ \hline
            k = 155 & -0.2040 & -0.2068 & 0.0640 & 0.3090 & 0.5280 & 0.2569 \\ \hline
            k = 156 & 1.1206 & -0.6124 & 0.7021 & 1.1042 & 0.5038 & -0.9468 \\
\end{tabular*}
		}
	\end{table}
	
	\begin{table}[tbp]
	\centering
	    \scalebox{0.99}{
		\begin{tabular*}{13cm}{@{\extracolsep{\fill}}lcccccccccccc}
            \\
			\hline\hline
			\textbf{$U_1$} & $n=1(x)$ & $n=2(x)$ & $n=3(x)$ & $n=1(y)$ & $n=2(y)$ & $n=3(y)$ \\ \hline
            k = 201 & -1.6040 & -0.8448 & 2.5724 & -1.5631 & 1.0153 & -0.7850 \\ \hline
            k = 202 & 0.1850 & 0.1846 & -0.4303 & 0.0155 & -0.0768 & -0.3154 \\ \hline
            k = 203 & 1.3274 & -0.1745 & 0.9304 & -0.2085 & 0.1299 & 0.2376 \\ \hline
            k = 204 & 1.7737 & -0.1940 & -1.3270 & 0.5072 & 0.5162 & -0.1313 \\ \hline
            k = 205 & 1.5486 & 0.7277 & -0.7744 & 3.0820 & -1.7889 & 0.4750 \\ \hline
            k = 206 & 1.9122 & -0.2521 & -0.3037 & 2.5793 & 0.1055 & -0.4412 \\ \hline
            k = 207 & 4.2434 & 1.2512 & -1.2009 & 3.0575 & -0.4410 & 1.1753 \\ \hline
            k = 208 & 3.4602 & -2.0269 & -0.5980 & 3.2769 & -3.5424 & -0.0797 \\ \hline
            k = 209 & 4.9201 & -0.0843 & 0.1663 & 4.9350 & -7.5115 & 0.3765 \\ \hline
            k = 210 & 3.0609 & -0.5940 & 1.9197 & 6.1301 & -13.2567 & -0.7216 \\ \hline
			\\ 
		\end{tabular*}
		}
	\end{table}
	
	\begin{table}[ht]
	\centering
		\caption{The magnetic field $\{h^x_{n,k}\}$ and $\{h^y_{n,k}\}$ to realize the $U_2$. We take the total time $T=2.1$, $K=210$, and thus the Trotter step $\tau=2.1/210$.} \label{tab-tabel}
		\scalebox{0.99}{
		\begin{tabular*}{13cm}{@{\extracolsep{\fill}}lcccccc}
            \\
			\hline\hline
			\textbf{$U_2$} & $n=1(x)$ & $n=2(x)$ & $n=2(x)$ & $n=1(y)$ & $n=2(y)$ & $n=2(y)$ \\ \hline
            k = 1 & -0.9321 & 2.2444 & 5.0748 & 15.8218 & 0.0510 & -2.0459 \\ \hline
            k = 2 & -1.2790 & -0.0636 & 4.4257 & 7.2754 & -0.4515 & -1.9556 \\ \hline
            k = 3 & 1.2962 & -0.2390 & 1.7553 & 2.2986 & 0.0086 & -0.8688 \\ \hline
            k = 4 & -0.9038 & 0.6719 & 2.7338 & 0.5179 & -0.7817 & -1.8169 \\ \hline
            k = 5 & 0.3002 & 0.2591 & 0.5652 & -1.0982 & -1.1519 & -0.1771 \\ \hline
            k = 6 & -0.0593 & -1.9728 & 0.6902 & -0.3732 & 0.5104 & 0.3538 \\
\end{tabular*}
		}
	\end{table}
	
	\begin{table}[tbp]
	\centering
	    \scalebox{0.99}{
		\begin{tabular*}{13cm}{@{\extracolsep{\fill}}lcccccccccccc}
            \\
			\hline\hline
			\textbf{$U_2$} & $n=1(x)$ & $n=2(x)$ & $n=3(x)$ & $n=1(y)$ & $n=2(y)$ & $n=3(y)$ \\ \hline
            k = 51 & 0.8638 & 0.1006 & -1.1436 & 0.1981 & -0.9254 & 0.2057 \\ \hline
            k = 52 & -0.9101 & 0.0996 & -0.5079 & -0.7331 & -0.0426 & 0.0845 \\ \hline
            k = 53 & -0.0080 & 0.0029 & 1.3779 & -0.0895 & 0.8093 & -0.6866 \\ \hline
            k = 54 & -0.4507 & 0.2945 & 0.0438 & -0.9654 & 1.3726 & -0.7635 \\ \hline
            k = 55 & 0.8989 & -0.3776 & -0.9052 & -0.1659 & -0.3638 & -0.3753 \\ \hline
            k = 56 & -0.5178 & 0.1098 & 0.5428 & 0.4260 & 0.1111 & 1.1257 \\
\end{tabular*}
		}
	\end{table}
	
	\begin{table}[tbp]
	\centering
	    \scalebox{0.99}{
		\begin{tabular*}{13cm}{@{\extracolsep{\fill}}lcccccccccccc}
            \\
			\hline\hline
			\textbf{$U_2$} & $n=1(x)$ & $n=2(x)$ & $n=3(x)$ & $n=1(y)$ & $n=2(y)$ & $n=3(y)$ \\ \hline
            k = 101 & 0.6057 & -1.0131 & -0.7978 & 1.2223 & -0.5644 & 0.1666 \\ \hline
            k = 102 & -0.5434 & 1.2605 & 0.8576 & 1.4137 & 0.9446 & -0.1119 \\ \hline
            k = 103 & -0.6791 & -1.3416 & 1.3308 & -1.2702 & 0.3304 & 0.2832 \\ \hline
            k = 104 & -0.0135 & -1.2531 & -1.2727 & -1.7741 & -0.6272 & 0.1118 \\ \hline
            k = 105 & -0.1895 & 0.6120 & -0.9407 & 0.7862 & -0.0254 & -1.0076 \\ \hline
            k = 106 & 0.1406 & 0.7285 & 1.1331 & 1.0889 & 0.0320 & 0.1747 \\
\end{tabular*}
		}
	\end{table}
	
	\begin{table}[tbp]
	\centering
	    \scalebox{0.99}{
		\begin{tabular*}{13cm}{@{\extracolsep{\fill}}lcccccccccccc}
            \\
			\hline\hline
			\textbf{$U_2$} & $n=1(x)$ & $n=2(x)$ & $n=3(x)$ & $n=1(y)$ & $n=2(y)$ & $n=3(y)$ \\ \hline
            k = 151 & 0.2527 & -0.7305 & -1.3446 & 0.7102 & -0.1333 & 1.1143 \\ \hline
            k = 152 & 1.6466 & -1.1449 & -0.2993 & 1.4097 & -1.1295 & -0.4325 \\ \hline
            k = 153 & 2.5597 & -0.6725 & -1.1527 & 0.9171 & -0.3002 & -0.6644 \\ \hline
            k = 154 & 1.3426 & -0.6200 & 0.1700 & -0.9980 & -0.4493 & 0.1966 \\ \hline
            k = 155 & 1.4351 & -1.1697 & -0.8807 & -0.6109 & 1.1026 & -0.2231 \\ \hline
            k = 156 & 0.3653 & 0.3289 & -0.5449 & -1.3587 & -0.6615 & -0.9834 \\
\end{tabular*}
		}
	\end{table}
	
	\begin{table}[tbp]
	\centering
	    \scalebox{0.99}{
		\begin{tabular*}{13cm}{@{\extracolsep{\fill}}lcccccccccccc}
            \\
			\hline\hline
			\textbf{$U_2$} & $n=1(x)$ & $n=2(x)$ & $n=3(x)$ & $n=1(y)$ & $n=2(y)$ & $n=3(y)$ \\ \hline
            k = 201 & -3.3502 & 0.8609 & -1.0468 & -2.2938 & 0.1745 & -0.7520 \\ \hline
            k = 202 & -1.8336 & -0.7190 & -0.1131 & -1.1447 & -0.0502 & 0.3246 \\ \hline
            k = 203 & -4.3332 & -0.0054 & -0.4384 & -2.0500 & -1.8797 & 0.9046 \\ \hline
            k = 204 & -4.4136 & 0.2021 & 0.1994 & -2.1387 & -0.3884 & 0.9044 \\ \hline
            k = 205 & -5.1375 & 1.0719 & -0.2773 & -3.3513 & -1.1329 & 0.1719 \\ \hline
            k = 206 & -5.1783 & 0.2201 & -0.5733 & -1.9876 & 0.3098 & -0.0315 \\ \hline
            k = 207 & -6.0091 & 1.2966 & 0.1418 & -2.5188 & 0.8608 & -1.7112 \\ \hline
            k = 208 & -5.9253 & 0.1547 & -0.4121 & -3.6755 & 1.1350 & -0.8394 \\ \hline
            k = 209 & -3.0352 & -0.5561 & 0.6317 & -3.1084 & -0.0399 & 1.3061 \\ \hline
            k = 210 & -5.7163 & -1.0228 & 0.7422 & -2.5126 & 0.3025 & -1.4748 \\ \hline
			\\ 
		\end{tabular*}
		}
	\end{table}
	
	\begin{table}[ht]
	\centering
		\caption{The magnetic field $\{h^x_{n,k}\}$ and $\{h^y_{n,k}\}$ to realize the $U_3$. We take the total time $T=1.4$, $K=140$, and thus the Trotter step $\tau=1.4/140$.} \label{tab-tabel}
		\scalebox{0.99}{
		\begin{tabular*}{13cm}{@{\extracolsep{\fill}}lcccccc}
            \\
			\hline\hline
			\textbf{$U_3$} & $n=1(x)$ & $n=2(x)$ & $n=1(y)$ & $n=2(y)$  \\ \hline
            k = 1 & -0.3013 & 0.6898 & 2.0804 & 2.3129 \\ \hline
            k = 2 & 1.9663 & -1.6758 & 2.8192 & 2.8023 \\ \hline
            k = 3 & 0.5549 & -0.9799 & 6.6289 & 5.8073 \\ \hline
            k = 4 & -3.3017 & 2.6915 & 3.8890 & 3.6518 \\ \hline
            k = 5 & -2.2577 & 2.0026 & 3.5685 & 3.9652 \\ \hline
            k = 6 & -1.0841 & 0.8933 & 2.7589 & 3.2097 \\
\end{tabular*}
		}
	\end{table}
	
	\begin{table}[tbp]
	\centering
	    \scalebox{0.99}{
		\begin{tabular*}{13cm}{@{\extracolsep{\fill}}lcccccccccccc}
            \\
			\hline\hline
			\textbf{$U_3$} & $n=1(x)$ & $n=2(x)$ & $n=1(y)$ & $n=2(y)$  \\ \hline
            k = 51 & -1.6813 & 0.2456 & 1.5494 & -2.3641 \\ \hline
            k = 52 & -3.4904 & -0.6832 & 2.2672 & -3.5932 \\ \hline
            k = 53 & -6.8864 & -3.4631 & 0.5759 & -5.3982 \\ \hline
            k = 54 & -4.3939 & -3.3412 & -1.1810 & -2.7840 \\ \hline
            k = 55 & -5.1608 & -3.6406 & -0.0280 & -3.1507 \\ \hline
            k = 56 & -9.1840 & -8.3819 & -2.9732 & -3.5767 \\
\end{tabular*}
		}
	\end{table}

    \begin{table}[tbp]
	\centering
	    \scalebox{0.99}{
		\begin{tabular*}{13cm}{@{\extracolsep{\fill}}lcccccccccccc}
            \\
			\hline\hline
			\textbf{$U_3$} & $n=1(x)$ & $n=2(x)$ & $n=1(y)$ & $n=2(y)$  \\ \hline
            k = 101 & -0.7701 & -1.5993 & 1.8172 & -0.4466 \\ \hline
            k = 102 & 1.1877 & 0.1617 & 0.1401 & -1.7052 \\ \hline
            k = 103 & 3.3483 & 2.8258 & -1.3492 & -2.6478 \\ \hline
            k = 104 & 2.7608 & 3.2752 & -2.0487 & -1.3747 \\ \hline
            k = 105 & 2.4689 & 2.5301 & -1.3436 & -1.3255 \\ \hline
            k = 106 & 0.7241 & 1.7005 & -1.5418 & 0.0607 \\
\end{tabular*}
		}
	\end{table}
	
	\begin{table}[ht]
	\centering
		\caption{The magnetic field $\{h^x_{n,k}\}$ and $\{h^y_{n,k}\}$ to realize the $U_4$. We take the total time $T=2.4$, $K=240$, and thus the Trotter step $\tau=2.4/240$.} \label{tab-tabel}
		\scalebox{0.99}{
		\begin{tabular*}{13cm}{@{\extracolsep{\fill}}lcccccc}
            \\
			\hline\hline
			\textbf{$U_4$} & $n=1(x)$ & $n=2(x)$ & $n=2(x)$ & $n=1(y)$ & $n=2(y)$ & $n=2(y)$ \\ \hline
            k = 1 & -3.1399 & -1.3311 & -0.3132 & 0.4410 & -0.5488 & 13.2111 \\ \hline
            k = 2 & -0.5521 & -0.2380 & -1.1714 & 1.8920 & -1.5596 & 7.4406 \\ \hline
            k = 3 & -0.1767 & -0.4128 & 0.7828 & 0.6015 & -0.7912 & 5.3326 \\ \hline
            k = 4 & 0.2408 & 0.3326 & -0.3585 & -0.6166 & 0.3082 & 4.1461 \\ \hline
            k = 5 & -0.4013 & 0.4130 & 0.4504 & 0.6606 & -0.0870 & 0.5155 \\ \hline
            k = 6 & 0.5652 & -0.1958 & 0.1217 & -0.2205 & -0.0274 & 1.8436 \\
\end{tabular*}
		}
	\end{table}
	
	\begin{table}[tbp]
	\centering
	    \scalebox{0.99}{
		\begin{tabular*}{13cm}{@{\extracolsep{\fill}}lcccccccccccc}
            \\
			\hline\hline
			\textbf{$U_4$} & $n=1(x)$ & $n=2(x)$ & $n=3(x)$ & $n=1(y)$ & $n=2(y)$ & $n=3(y)$ \\ \hline
            k = 51 & -1.5698 & 0.8711 & -1.1907 & 1.7242 & 1.5744 & 0.5597 \\ \hline
            k = 52 & -1.1742 & 0.9349 & 1.1554 & -0.0185 & 2.0398 & -0.5263 \\ \hline
            k = 53 & -2.1322 & -0.1598 & 0.6798 & 1.4861 & 2.5103 & -1.6766 \\ \hline
            k = 54 & -1.1913 & 1.1856 & 2.5809 & 1.4748 & 2.2867 & -2.7907 \\ \hline
            k = 55 & -1.9647 & 0.7391 & 2.9639 & 0.3857 & 1.5100 & -2.1667 \\ \hline
            k = 56 & 0.2022 & 0.6071 & 3.3680 & 1.5482 & 1.5099 & -4.6117 \\
\end{tabular*}
		}
	\end{table}
	
	\begin{table}[tbp]
	\centering
	    \scalebox{0.99}{
		\begin{tabular*}{13cm}{@{\extracolsep{\fill}}lcccccccccccc}
            \\
			\hline\hline
			\textbf{$U_4$} & $n=1(x)$ & $n=2(x)$ & $n=3(x)$ & $n=1(y)$ & $n=2(y)$ & $n=3(y)$ \\ \hline
            k = 101 & -1.9844 & -3.1773 & 1.2756 & -1.7699 & 0.1711 & 1.0828 \\ \hline
            k = 102 & -0.3046 & -1.1774 & -0.6470 & -1.1303 & 0.5014 & -0.9075 \\ \hline
            k = 103 & -2.4655 & -1.3024 & 0.9717 & -2.1335 & 2.1251 & 2.0064 \\ \hline
            k = 104 & 0.0023 & -1.4732 & 0.9782 & -1.6488 & -0.4384 & 0.8949 \\ \hline
            k = 105 & -0.9081 & -0.9320 & -0.4786 & -0.2323 & 0.0264 & -1.0107 \\ \hline
            k = 106 & -0.4276 & -0.3397 & 1.8277 & -0.7637 & -0.7670 & -1.3961 \\
\end{tabular*}
		}
	\end{table}
	
	\begin{table}[tbp]
	\centering
	    \scalebox{0.99}{
		\begin{tabular*}{13cm}{@{\extracolsep{\fill}}lcccccccccccc}
            \\
			\hline\hline
			\textbf{$U_4$} & $n=1(x)$ & $n=2(x)$ & $n=3(x)$ & $n=1(y)$ & $n=2(y)$ & $n=3(y)$ \\ \hline
            k = 151 & -0.0381 & -0.2964 & 1.1658 & -0.7654 & 0.6615 & -0.2786 \\ \hline
            k = 152 & -0.4272 & 1.7936 & -0.9579 & -1.5698 & -1.3101 & 0.2235 \\ \hline
            k = 153 & 1.1254 & 0.0933 & -1.5330 & -2.5213 & 1.1144 & -0.0353 \\ \hline
            k = 154 & 1.2585 & 1.9223 & -0.8864 & -1.2513 & -0.4482 & -1.4215 \\ \hline
            k = 155 & -0.1030 & 0.1198 & 0.2463 & -1.8194 & -1.3198 & -0.4503 \\ \hline
            k = 156 & 3.1583 & 1.8197 & -0.6107 & -1.3098 & -0.4684 & 0.4575 \\
\end{tabular*}
		}
	\end{table}
	
	\begin{table}[tbp]
	\centering
	    \scalebox{0.99}{
		\begin{tabular*}{13cm}{@{\extracolsep{\fill}}lcccccccccccc}
            \\
			\hline\hline
			\textbf{$U_4$} & $n=1(x)$ & $n=2(x)$ & $n=3(x)$ & $n=1(y)$ & $n=2(y)$ & $n=3(y)$ \\ \hline
            k = 201 & 0.5653 & -0.5405 & -0.7020 & 0.2396 & 1.0270 & 0.3285 \\ \hline
            k = 202 & -1.5339 & 0.9237 & -0.3006 & 0.3722 & 0.2643 & 1.2840 \\ \hline
            k = 203 & -0.1794 & -0.9906 & -1.4394 & -0.5788 & 1.3299 & 0.6026 \\ \hline
            k = 204 & 0.9270 & -0.0349 & 1.4508 & -0.6796 & -0.0544 & 0.6770 \\ \hline
            k = 205 & 0.1871 & 1.0170 & -1.3885 & 0.4854 & -0.2825 & -0.0113 \\ \hline
            k = 206 & 0.5602 & -0.9537 & 0.4518 & 0.3764 & -0.0677 & -0.7876 \\
\end{tabular*}
		}
	\end{table}
	
	\begin{table}[ht]
	\centering
		\caption{The magnetic field $\{h^x_{n,k}\}$ and $\{h^y_{n,k}\}$ to realize the $U_5$. We take the total time $T=1.5$, $K=150$, and thus the Trotter step $\tau=1.5/150$.} \label{tab-tabel}
		\scalebox{0.99}{
		\begin{tabular*}{13cm}{@{\extracolsep{\fill}}lcccccc}
            \\
			\hline\hline
			\textbf{$U_5$} & $n=1(x)$ & $n=2(x)$ & $n=1(y)$ & $n=2(y)$  \\ \hline
            k = 1 & -0.2526 & 0.1246 & 3.7404 & 3.0209 \\ \hline
            k = 2 & 1.6681 & 0.6400 & 2.7494 & 3.0677 \\ \hline
            k = 3 & 2.4961 & -2.2851 & 2.7072 & 2.1781 \\ \hline
            k = 4 & 2.9811 & -1.7380 & 1.9319 & 3.8018 \\ \hline
            k = 5 & 1.9753 & -1.3127 & 3.7385 & 3.9821 \\ \hline
            k = 6 & 2.4912 & -0.8143 & 2.4790 & 3.1630 \\
\end{tabular*}
		}
	\end{table}
	
	\begin{table}[tbp]
	\centering
	    \scalebox{0.99}{
		\begin{tabular*}{13cm}{@{\extracolsep{\fill}}lcccccccccccc}
            \\
			\hline\hline
			\textbf{$U_5$} & $n=1(x)$ & $n=2(x)$ & $n=1(y)$ & $n=2(y)$  \\ \hline
            k = 51 & 2.5880 & -3.7934 & 1.9601 & 0.7360 \\ \hline
            k = 52 & 2.8453 & -3.0569 & 2.6682 & 0.7348 \\ \hline
            k = 53 & 1.6256 & -1.3941 & 0.0388 & 0.2349 \\ \hline
            k = 54 & 1.0472 & -0.0278 & -0.5968 & -0.8466 \\ \hline
            k = 55 & -0.2945 & -0.7209 & 0.4633 & 0.1301 \\ \hline
            k = 56 & 1.9436 & -1.2473 & 0.5889 & -1.4228 \\
\end{tabular*}
		}
	\end{table}

    \begin{table}[tbp]
	\centering
	    \scalebox{0.99}{
		\begin{tabular*}{13cm}{@{\extracolsep{\fill}}lcccccccccccc}
            \\
			\hline\hline
			\textbf{$U_5$} & $n=1(x)$ & $n=2(x)$ & $n=1(y)$ & $n=2(y)$  \\ \hline
            k = 101 & 1.2317 & 2.9320 & -1.5134 & -1.2433 \\ \hline
            k = 102 & 1.3720 & 0.0728 & 0.9427 & -2.0653 \\ \hline
            k = 103 & 1.6216 & 0.3880 & -0.3022 & -1.3144 \\ \hline
            k = 104 & 0.5013 & -0.2014 & -0.3159 & 0.1511 \\ \hline
            k = 105 & 0.3475 & 2.1549 & -1.5947 & -0.0111 \\ \hline
            k = 106 & -0.4067 & -1.3875 & 1.4739 & -1.4629 \\
\end{tabular*}
		}
	\end{table}
	
	\begin{table}[ht]
	\centering
		\caption{The magnetic field $\{h^x_{n,k}\}$ and $\{h^y_{n,k}\}$ to realize the $U_6$. We take the total time $T=2.4$, $K=240$, and thus the Trotter step $\tau=2.4/240$.} \label{tab-tabel}
		\scalebox{0.99}{
		\begin{tabular*}{13cm}{@{\extracolsep{\fill}}lcccccc}
            \\
			\hline\hline
			\textbf{$U_6$} & $n=1(x)$ & $n=2(x)$ & $n=2(x)$ & $n=1(y)$ & $n=2(y)$ & $n=2(y)$ \\ \hline
            k = 1 & -0.2497 & 0.5100 & 4.6057 & 0.1791 & -1.0998 & 2.6481 \\ \hline
            k = 2 & 0.0249 & 0.7050 & 1.8959 & 0.4812 & -0.1437 & 2.2382 \\ \hline
            k = 3 & -0.0737 & 0.0018 & 3.2687 & -0.5180 & 0.3367 & 4.3415 \\ \hline
            k = 4 & 0.3004 & 0.6355 & 2.2547 & -0.5387 & 0.6146 & 4.6550 \\ \hline
            k = 5 & -0.3977 & 0.0744 & 0.9662 & 0.3782 & -0.7550 & 1.1707 \\ \hline
            k = 6 & -0.1321 & -0.8564 & 3.0828 & 0.2397 & 0.2449 & 1.2369 \\
\end{tabular*}
		}
	\end{table}
	
	\begin{table}[tbp]
	\centering
	    \scalebox{0.99}{
		\begin{tabular*}{13cm}{@{\extracolsep{\fill}}lcccccccccccc}
            \\
			\hline\hline
			\textbf{$U_6$} & $n=1(x)$ & $n=2(x)$ & $n=3(x)$ & $n=1(y)$ & $n=2(y)$ & $n=3(y)$ \\ \hline
            k = 51 & 0.2293 & -0.1974 & 1.3549 & 0.4354 & 0.3913 & -2.0596 \\ \hline
            k = 52 & 0.1708 & -0.2360 & 1.6599 & -0.2171 & 0.0145 & -2.9738 \\ \hline
            k = 53 & 0.9497 & -0.6061 & 0.0885 & 0.2058 & 1.0790 & -3.2615 \\ \hline
            k = 54 & 0.9410 & -0.3143 & 0.9997 & 0.4361 & 0.3976 & -1.2407 \\ \hline
            k = 55 & 0.0385 & -0.8826 & 0.2962 & -0.1107 & 0.3532 & -2.6491 \\ \hline
            k = 56 & 0.8990 & -0.1062 & -0.5688 & 0.4652 & 0.5910 & -4.0023 \\
\end{tabular*}
		}
	\end{table}
	
	\begin{table}[tbp]
	\centering
	    \scalebox{0.99}{
		\begin{tabular*}{13cm}{@{\extracolsep{\fill}}lcccccccccccc}
            \\
			\hline\hline
			\textbf{$U_6$} & $n=1(x)$ & $n=2(x)$ & $n=3(x)$ & $n=1(y)$ & $n=2(y)$ & $n=3(y)$ \\ \hline
            k = 101 & 0.6342 & -0.4127 & 0.3592 & -0.1613 & 0.9504 & -0.4756 \\ \hline
            k = 102 & 0.9622 & -1.3255 & 0.2260 & 1.2500 & 0.2236 & 0.2948 \\ \hline
            k = 103 & 1.1919 & -0.2715 & -0.1248 & 0.4849 & 1.3527 & -0.5681 \\ \hline
            k = 104 & 1.0399 & -0.6070 & 0.4787 & 0.5657 & 0.1895 & -0.9628 \\ \hline
            k = 105 & 1.1831 & -1.1931 & -0.6240 & 0.6814 & 1.7484 & -0.0850 \\ \hline
            k = 106 & 1.4211 & 0.0566 & 1.1138 & -0.6054 & 0.7383 & -0.8036 \\
\end{tabular*}
		}
	\end{table}
	
	\begin{table}[tbp]
	\centering
	    \scalebox{0.99}{
		\begin{tabular*}{13cm}{@{\extracolsep{\fill}}lcccccccccccc}
            \\
			\hline\hline
			\textbf{$U_6$} & $n=1(x)$ & $n=2(x)$ & $n=3(x)$ & $n=1(y)$ & $n=2(y)$ & $n=3(y)$ \\ \hline
            k = 151 & 2.5441 & -0.1285 & -0.5492 & 1.9298 & 1.0187 & -0.2967 \\ \hline
            k = 152 & -0.5263 & -0.4344 & -0.1148 & 1.5449 & -0.0968 & -0.3728 \\ \hline
            k = 153 & 2.5069 & -1.9371 & 1.8807 & 3.3706 & 1.0012 & -0.9541 \\ \hline
            k = 154 & 2.8177 & -1.2702 & -0.2051 & 3.5884 & 1.7461 & -1.7164 \\ \hline
            k = 155 & 3.5532 & -2.4746 & 0.2541 & 5.4222 & 2.0584 & -1.2955 \\ \hline
            k = 156 & 3.6865 & -3.0133 & 1.6482 & 4.5209 & 1.8216 & -2.0246 \\
\end{tabular*}
		}
	\end{table}
	
	\begin{table}[tbp]
	\centering
	    \scalebox{0.99}{
		\begin{tabular*}{13cm}{@{\extracolsep{\fill}}lcccccccccccc}
            \\
			\hline\hline
			\textbf{$U_6$} & $n=1(x)$ & $n=2(x)$ & $n=3(x)$ & $n=1(y)$ & $n=2(y)$ & $n=3(y)$ \\ \hline
            k = 201 & -1.0091 & 0.0717 & 0.1447 & -2.3102 & -0.9485 & 0.3595 \\ \hline
            k = 202 & 0.7319 & -0.7980 & 1.9473 & -1.6241 & -0.8940 & 0.0608 \\ \hline
            k = 203 & -0.6019 & -0.2369 & -1.3414 & -1.8735 & 0.6588 & 0.1589 \\ \hline
            k = 204 & 0.7618 & 0.6465 & 0.6701 & -2.0000 & -0.1725 & 0.2333 \\ \hline
            k = 205 & -0.9652 & -1.0499 & 0.2322 & -1.3188 & -0.0966 & -0.5307 \\ \hline
            k = 206 & -0.0716 & 0.0247 & -0.1714 & -1.4013 & 0.1459 & 0.2868 \\
\end{tabular*}
		}
	\end{table}	
	
			\begin{table}[ht]
	\centering
		\caption{The magnetic field $\{h^x_{n,k}\}$ and $\{h^y_{n,k}\}$ to realize the $U_7$. We take the total time $T=1.5$, $K=150$, and thus the Trotter step $\tau=1.5/150$.} \label{tab-tabel}
		\scalebox{0.99}{
		\begin{tabular*}{13cm}{@{\extracolsep{\fill}}lcccccc}
            \\
			\hline\hline
			\textbf{$U_7$} & $n=1(x)$ & $n=2(x)$ & $n=1(y)$ & $n=2(y)$  \\ \hline
            k = 1 & -0.3705 & 0.7404 & -2.2725 & 1.9695 \\ \hline
            k = 2 & -0.5166 & 0.9703 & -2.9934 & 3.2378 \\ \hline
            k = 3 & -0.8282 & 0.3253 & -4.6689 & 5.5886 \\ \hline
            k = 4 & -3.4511 & -3.9501 & -2.7337 & 4.6895 \\ \hline
            k = 5 & -2.9346 & -4.4916 & 1.2076 & -0.2509 \\ \hline
            k = 6 & -3.5153 & -4.2509 & 2.1102 & -0.6278 \\
\end{tabular*}
		}
	\end{table}
	
	\begin{table}[tbp]
	\centering
	    \scalebox{0.99}{
		\begin{tabular*}{13cm}{@{\extracolsep{\fill}}lcccccccccccc}
            \\
			\hline\hline
			\textbf{$U_7$} & $n=1(x)$ & $n=2(x)$ & $n=1(y)$ & $n=2(y)$  \\ \hline
            k = 51 & -0.6962 & 3.0570 & -2.6846 & 0.1486 \\ \hline
            k = 52 & -2.3608 & 4.4487 & -3.4551 & 1.3807 \\ \hline
            k = 53 & -4.3060 & 6.0901 & -4.6075 & 3.9650 \\ \hline
            k = 54 & -4.9172 & 3.7296 & -2.5484 & 6.2037 \\ \hline
            k = 55 & -4.4166 & -1.2446 & 2.4751 & 6.3218 \\ \hline
            k = 56 & -4.3375 & -1.8877 & 2.9552 & 4.8968 \\
\end{tabular*}
		}
	\end{table}

    \begin{table}[tbp]
	\centering
	    \scalebox{0.99}{
		\begin{tabular*}{13cm}{@{\extracolsep{\fill}}lcccccccccccc}
            \\
			\hline\hline
			\textbf{$U_7$} & $n=1(x)$ & $n=2(x)$ & $n=1(y)$ & $n=2(y)$  \\ \hline
            k = 101 & -4.7892 & 3.7055 & 0.7347 & -4.9249 \\ \hline
            k = 102 & -4.5129 & 6.3534 & 3.6925 & -2.8425 \\ \hline
            k = 103 & -1.6407 & 3.2257 & 1.9700 & 0.7434 \\ \hline
            k = 104 & 0.9135 & 0.1234 & -0.3691 & 1.1464 \\ \hline
            k = 105 & 2.5218 & 2.4065 & 2.2181 & 2.4364 \\ \hline
            k = 106 & 4.2815 & 1.0633 & 2.1705 & 5.5850 \\
\end{tabular*}
		}
	\end{table}
	
	\begin{table}[ht]
	\centering
		\caption{The magnetic field $\{h^x_{n,k}\}$ and $\{h^y_{n,k}\}$ to realize the $U_8$. We take the total time $T=2.4$, $K=240$, and thus the Trotter step $\tau=2.4/240$.} \label{table-9}
		\scalebox{0.99}{
		\begin{tabular*}{13cm}{@{\extracolsep{\fill}}lcccccc}
            \\
			\hline\hline
			\textbf{$U_8$} & $n=1(x)$ & $n=2(x)$ & $n=2(x)$ & $n=1(y)$ & $n=2(y)$ & $n=2(y)$ \\ \hline
            k = 1 & -0.1964 & -0.1354 & 0.5574 & 0.6408 & -0.2138 & 0.4839 \\ \hline
            k = 2 & 0.0325 & 0.1663 & 1.4338 & -0.4663 & 0.0921 & 0.0175 \\ \hline
            k = 3 & 0.3204 & 0.4952 & -0.4223 & 0.2553 & -0.1883 & -3.0474 \\ \hline
            k = 4 & -0.5582 & -0.2317 & -1.5425 & -0.2554 & -0.1940 & 0.5426 \\ \hline
            k = 5 & 0.1921 & -0.1450 & -1.8036 & 0.2723 & 0.0540 & 0.8655 \\ \hline
            k = 6 & 0.2638 & -0.2091 & 1.2852 & -0.4073 & 0.1310 & 1.4299 \\
\end{tabular*}
		}
	\end{table}
	
	\begin{table}[tbp]
	\centering
	    \scalebox{0.99}{
		\begin{tabular*}{13cm}{@{\extracolsep{\fill}}lcccccccccccc}
            \\
			\hline\hline
			\textbf{$U_8$} & $n=1(x)$ & $n=2(x)$ & $n=3(x)$ & $n=1(y)$ & $n=2(y)$ & $n=3(y)$ \\ \hline
            k = 51 & -0.7311 & 0.9778 & -2.1063 & 1.5264 & -1.5150 & 1.4853 \\ \hline
            k = 52 & -1.0637 & 0.9191 & -2.4300 & 1.3508 & -1.5941 & 2.5860 \\ \hline
            k = 53 & -0.9906 & 1.8992 & -5.4433 & 2.4631 & -2.3271 & 4.7369 \\ \hline
            k = 54 & -1.4883 & 1.3989 & -4.7217 & 1.7757 & -2.1774 & 5.8143 \\ \hline
            k = 55 & -0.5910 & 1.2083 & -3.6907 & 1.0375 & -1.1325 & 6.6304 \\ \hline
            k = 56 & -0.3842 & 0.4904 & -2.0815 & 0.6741 & -1.0001 & 7.0988 \\
\end{tabular*}
		}
	\end{table}
	
	\begin{table}[tbp]
	\centering
	    \scalebox{0.99}{
		\begin{tabular*}{13cm}{@{\extracolsep{\fill}}lcccccccccccc}
            \\
			\hline\hline
			\textbf{$U_8$} & $n=1(x)$ & $n=2(x)$ & $n=3(x)$ & $n=1(y)$ & $n=2(y)$ & $n=3(y)$ \\ \hline
            k = 101 & 0.5999 & 0.3717 & 0.1938 & -0.6864 & 0.2490 & -0.1972 \\ \hline
            k = 102 & -0.9259 & -0.8905 & -0.3828 & 0.1348 & -0.0080 & 0.4886 \\ \hline
            k = 103 & -0.5460 & -0.8173 & 0.7303 & 0.3933 & 0.1030 & 0.2454 \\ \hline
            k = 104 & -1.4416 & -0.8336 & 0.4176 & -0.9764 & 0.4933 & 0.0653 \\ \hline
            k = 105 & -0.7658 & -0.9364 & 0.3241 & 0.0540 & 0.3319 & 0.0873 \\ \hline
            k = 106 & -0.9889 & -1.0460 & 0.0772 & -0.2409 & -0.2708 & 1.3143 \\
\end{tabular*}
		}
	\end{table}
	
	\begin{table}[tbp]
	\centering
	    \scalebox{0.99}{
		\begin{tabular*}{13cm}{@{\extracolsep{\fill}}lcccccccccccc}
            \\
			\hline\hline
			\textbf{$U_8$} & $n=1(x)$ & $n=2(x)$ & $n=3(x)$ & $n=1(y)$ & $n=2(y)$ & $n=3(y)$ \\ \hline
            k = 151 & 0.0808 & 0.9299 & 0.5641 & 0.0420 & 0.5412 & -0.9917 \\ \hline
            k = 152 & 0.2886 & 0.4480 & -0.0449 & -0.0209 & 0.0806 & -0.0808 \\ \hline
            k = 153 & 1.1902 & 0.5130 & -0.8305 & 0.2831 & -0.4322 & -0.1359 \\ \hline
            k = 154 & 0.0006 & 0.7637 & -0.3311 & 0.3710 & -0.2660 & -0.4119 \\ \hline
            k = 155 & -0.6007 & 0.2732 & -0.4268 & 0.3711 & 0.0529 & -0.2235 \\ \hline
            k = 156 & -0.8136 & 1.1444 & -0.1257 & -0.1407 & 0.0181 & -0.7677 \\
\end{tabular*}
		}
	\end{table}
	
	\begin{table}[tbp]
	\centering
	    \scalebox{0.99}{
		\begin{tabular*}{13cm}{@{\extracolsep{\fill}}lcccccccccccc}
            \\
			\hline\hline
			\textbf{$U_8$} & $n=1(x)$ & $n=2(x)$ & $n=3(x)$ & $n=1(y)$ & $n=2(y)$ & $n=3(y)$ \\ \hline
            k = 201 & 0.1851 & -0.3027 & -0.2313 & -0.6224 & -0.2308 & 0.4237 \\ \hline
            k = 202 & 0.3373 & -0.1564 & -0.1595 & 0.4800 & -0.4786 & 0.4838 \\ \hline
            k = 203 & -0.4025 & -0.1951 & -0.2184 & 0.1315 & -0.1005 & 0.2085 \\ \hline
            k = 204 & 0.1924 & 0.1927 & -0.0123 & -0.2683 & -0.0073 & -0.2202 \\ \hline
            k = 205 & -0.6061 & -0.0088 & -0.0667 & -0.4358 & 0.0909 & -0.1600 \\ \hline
            k = 206 & 0.8832 & 0.0074 & 0.4528 & -0.1001 & 0.0337 & 0.1368 \\
\end{tabular*}
		}
	\end{table}

\end{document}